\documentclass[a4paper, 11pt]{article}
\usepackage{amsmath,amsthm,amssymb,mathrsfs,mathtools,graphicx,color,yhmath}
\usepackage[titletoc,title]{appendix}
\usepackage[normalem]{ulem}

\usepackage{bm}
\usepackage{here}
\usepackage{bbm}
\usepackage[hang,small,bf]{caption}
\usepackage[subrefformat=parens]{subcaption}
\newcommand{\nc}{\newcommand}
\nc{\rnc}{\renewcommand}
\nc{\nn}{\nonumber}
\nc{\ms}{\mathsf}
\nc{\mr}{\mathrm}
\nc{\mf}{\mathfrak}

\nc{\del}{{\partial}}
\rnc{\Im}{{\textrm{Im}\,}}
\rnc{\Re}{{\textrm{Re}\,}}

\nc{\bra}{\langle}
\nc{\ket}{\rangle}
\nc{\dr}{\mathrm{dr}}
\nc{\for}{\textrm{for}}
\nc{\s}{\sigma}
\nc{\la}{\lambda}
\nc{\g}{\boldsymbol{\mathcal{G}}}
\nc{\n}{\tilde{n}}

\makeatletter
\newsavebox{\@brx}
\newcommand{\llangle}[1][]{\savebox{\@brx}{\(\m@th{#1\langle}\)}%
  \mathopen{\copy\@brx\mkern2mu\kern-0.9\wd\@brx\usebox{\@brx}}}
\newcommand{\rrangle}[1][]{\savebox{\@brx}{\(\m@th{#1\rangle}\)}%
  \mathclose{\copy\@brx\mkern2mu\kern-0.9\wd\@brx\usebox{\@brx}}}
\makeatother
\nc{\bbra}{\llangle}
\nc{\kket}{\rrangle}

\nc{\A}{\textrm{A}}
\nc{\B}{\textrm{B}}
\nc{\C}{\textrm{C}}
\nc{\D}{\textrm{D}}
\nc{\tcr}{\textcolor{red}}
\nc{\tcb}{\textcolor{blue}}

\DeclareMathOperator{\sgn}{sgn}

\theoremstyle{definition}

\numberwithin{equation}{section}

\begin{document}
\title{Long-range correlation and the spin conductivity in the XXZ chain from ballistic macroscopic fluctuation theory}

\author{
Shinya Ae \thanks{E-mail: 1221701@alumni.tus.ac.jp} 
\\\\
\textit{Tokyo Metropolitan Kasai Minami High School}\\
 \textit{Minami-Kasai 1-11-1, Edogawa-ku, Tokyo 134-8555, Japan} 
}

\date{}

\maketitle

\begin{abstract}
Based on the ballistic macroscopic fluctuation theory, the integration of the spin correlation function (spin conductivity) is analyzed for the spin-1/2 XXZ chain in the critical regime.
In the time when the magnetization of an infinite spin chain fluctuates from an initial state with a wavelength as long as the infinite length $N$, the equal-time two-point spin correlation function is scaled up to $O(1/N)$.
In the state where the ballistic spin transport decays at high temperature $T$, the diffusive transport remains on a large scale. We show that the spin conductivity is proportional to $1/T$ in the limit $T\to\infty$ and its high temperature proportionality constant diverges in the case where one-quasiparticle magnetization is infinitely large. This analysis informs that the superdiffusive spin transport is driven by the $1/N$-scaled long-range spin correlation and sheds a light on the dynamic scaling in spin transport at the isotropic point.
\end{abstract}

\section{Introduction} \label{Intro.}
The purpose of this paper is to analyze the fluctuating spin current in the XXZ chain which is driven by long-range spin correlation. We treat the model in the critical regime. For this analysis we use the ballistic macroscopic fluctuating theory (BMFT) \cite{BMFT 1, BMFT 2}, which is a new universal framework describing fluctuations and correlations in many-body systems.

At the thermodynamic equilibrium with nonzero temperatures, a finite correlation length of $\bra S^z_jS^z_k\ket$ is obtained for the spin operator $S^z_j$ using the quantum transfer matrices (QTMs) and their functional relations referred to as the $T$- and $Y$-systems \cite{KSS}, which results in an exponentially decaying $\bra S^z_jS^z_k\ket$ between two points on an infinite spin chain.
But a greater magnitude of long-range spin correlation unveils itself in non-equilibrium states through coarse-grained approximation per hydrodynamic fluid cell.
The BMFT reveals that when two or more conserved charges exist in a many-body system of interacting particles and the charge concentration fluctuates with a long wavelength $\ell$,
the equal-time charge correlation function is scaled up to the magnitude which is proportional to $1/\ell$ between two fluid cells far apart from each other \cite{Doyon25(2)}.    
Actually, this $1/\ell$-scaled long-range correlation has been verified by exactly calculating the equal-time two-point function of the particle density in the hard-rod gas \cite{BMFT 1, BMFT 2}. 
On the other hand the XXZ chain is a spin interaction system.
Since this is a Bethe ansatz integrable system, infinitely many conserved charges $Q_n$ exists from the logarithmic derivative of the transfer matrix $T(v)$ which commutes as $[T(v), T(v')]=0$ for different values of the spectral parameter $v, v' \in \mathbb{C}$ \footnote
{The $T(v)$ is given by the row transfer matrix $T_n^\mr{R}(v)$ in Ref.\,\cite{AS} as $T(v)=T_1^\mr{R}(v)$ and related to the auxiliary transfer matrix $T_A(u,v)$ in Ref.\,\cite{KSS} as
\begin{equation}
\left.\frac{\del}{\del u}T_A(u,0)\right|_{u=0}=\left.2\frac{d}{dv}\ln T(v)\right|_{v=0}.  \nn
\end{equation}
}:
\begin{equation}
Q_n=\left.\frac{d^{\,n}}{dv^n}\ln T(v)\right|_{v=0}.
\label{Qn}\end{equation}
In particular, $Q_1$ is the Hamiltonian and $Q_2$ is the energy current at zero magnetic field.
Thus, we will observe the $1/\ell$-scaled long-range correlations of conserved charges when the system evolves from such an initial magnetization concentration which is set up by and released from such an inclined magnetic field in the $z$ direction $h_0(\mr{x}/N)$ as
\begin{equation}
h_0(\mr{x}/N)
=
\begin{dcases}    
\left(1+\frac{2\mr{x}-Ld}{Nd}\right)h_0   \quad \for\;\;     \mr{x}\in \{\mr{x}_i\}_{i=1}^{N/2L} \\
\left(1-\frac{2\mr{x}-Ld}{Nd}\right)h_0   \quad \for\;\; \mr{x}\in \{\mr{x}_i\}_{i=N/2L+1}^{N/L}
                     \end{dcases};
\quad  \mr{x}_i=iLd-\frac{Nd}{2},
\label{m. field}\end{equation}
where $N\equiv 0\mod{2L}$ is the number of the lattice sites and $d$ is the lattice constant, which is set as $d=1$. The wavelength of the field and magnetization concentration is $\ell=N$.
$L$ is the length of the fluid cells. In an infinite spin chain ($N\to\infty$), $L$ is scaled as
\begin{equation}
\frac{L}{N} \to 0 \quad \textup{and} \quad L\to\infty,
\label{limits}\end{equation} 
which means that the fluid cells are sufficiently short so that the magnetization concentration is almost flat within each cell.
But they are so long that each cell reaches equilibrium locally with the field $h_0(\mr{x}/N)$ applied for a long time, and the magnetization concentration, or any other physical quantities in the respective cell can be approximated by the local thermal average at a temperature.

After cutting off the field $h_0(\mr{x}/N)$, the magnetization concentration starts to change.
In this time evolution, we anticipate an immediate effect of the $1/\ell$-scaled long-range correlation on charge currents.
To see that, let us define the conserved densities $\mr{q}_n(\mr{x},\mr{t})$ and their currents $\mr{j}_n(\mr{x},\mr{t})$, the former of which compose the conserved charges as $Q_n=L\sum_\mr{x}\mr{q}_n(\mr{x},\mr{t})$ \eqref{Qn} and both of which satisfy the hydrodynamic equations of motion
\begin{equation}
\del_\mr{t} \mr{q}_n(\mr{x},\mr{t})+\del_\mr{x}\mr{j}_n(\mr{x},\mr{t})=0.
\label{eq.o.m}\end{equation}
Furthermore, we rely on the equation of state 
$
\bbra\mr{j}_{n}(\mr{x},\mr{t})\kket=\mr{j}_{n}[\{\bbra \mr{q}_{n'}(\mr{x},\mr{t})\kket\}_{n'=1}^\infty]
$,
which is derived from a basic property of fluid systems that a current density of a conserved quantity is fully determined by some or all of the conserved densities. Here $\bbra\bullet \kket$ denotes a physical observation in the time evolution, which we approximate using the distribution functions of quasiparticles and holes in the sequel (see Eq. \eqref{relaxing s}).
By means of the equation of state, observing a flux $\bbra\mr{j}_{n}(\mr{x},\mr{t})\kket$ at a local fluid cell is a result of the variations of conserved densities in the same fluid cell but which may be driven by correlations with other fluid cells.

In fact, a remarkable discovery has been made in \cite{Hubner etc.} which shows that the diffusive flux of a conserved charge is described by the $1/\ell$-scaled equal-time two-point functions of conserved charges in the case of linear degenerate systems (see \cite{Doyon lec.} for lecture notes) to which integrable systems belong. 
This diffusion can not be treated in the frame of the Fick's law, a linear relation with the gradient of average densities, which is written as
$\bbra \mr{j}_n(\mr{x},\mr{t})\kket=-2^{-1}\sum_{m=1}^\infty \mathcal{D}_{nm}(\mr{x},\mr{t})\del_\mr{x}\bbra \mr{q}_m(\mr{x},\mr{t})\kket$ with the diffusion constant $\mathcal{D}_{nm}(\mr{x},\mr{t})$.
However, 
this diffusion is certainly connected with the current fluctuation of conserved charge.
To see this connection in the present system, let us introduce the spin density $\mr{q}_0(\mr{x},\mr{t})$ and its current $\mr{j}_0(\mr{x},\mr{t})$ and assume that all fluid cells have relaxed at a uniform temperature before removing the magnetic field.  
For then, the following relation is obtained in a long time after its removal as we show in Section \ref{Sec.s}:
\begin{align}
&\frac{\beta L}{\tau}\sum_{i=1}^{N/L} \mr{x}_i
                \bbra( \mr{q}_0(0,\mathcal{T})-\bbra\mr{q}_0(0,\mathcal{T})\kket)(\mr{q}_0(\mr{x}_i,\mathcal{T})-                  
                \bbra\mr{q}_0(\mr{x}_i,\mathcal{T})\kket)\kket  \nn\\
&=
\lim_{\mathcal{T}\to\infty}\frac{\beta L}{\mathcal{T}}\int_0^\mathcal{T} d\mr{t} \int_0^\mathcal{T} d\mr{t}' \sum_{i=1}^{N/L}   \bbra(\mr{j}_0(0,\mr{t}')-\bbra\mr{j}_0(0,\mr{t}')\kket)(\mr{j}_0(\mr{x}_i,\mr{t})-\bbra\mr{j}_0(\mr{x}_i,\mr{t})\kket)\kket    \nn\\
&=:\s(\beta).
\label{F.D.}\end{align}     
Here $\beta:=1/k_\mr{B}T$ is the inverse of the uniform temperature, $\mathcal{T}$ is the long time after the removal of the field, $\tau:=\mathcal{T}/N$ and $\s(\beta)$ is defined by the second expression of Eq.\,\eqref{F.D.}, which describes the spin conductivity. Actually, $\s(\beta)$ corresponds to the spin dc conductivity at the thermodynamic equilibrium (see Eq.\,\eqref{linear resp}).
We calculated the high temperature limit of the regular spin dc conductivity $\lim_{\beta\to0}\s^\mr{reg}(\beta)$ in \cite{Ae 2024}; we found that this quantity is proportional to $\beta$ in the limit $\beta\to0$ and the constant $\lim_{\beta\to0}\s^\mr{reg}(\beta)/\beta$ diverges in the case where one-particle magnetization is infinitely large.
In this case the spin transport is superdiffusive via the Einstein relation which connects the conductivity with the diffusion constant. 
In this paper, we inquire if the spin conductivity \eqref{F.D.} also diverges to see whether the superdiffusion is driven by the long-range spin correlation.

A good example of employing the BMFT has been provided in \cite{Yoshimura & Krajnik}, where this theory clarifies the origin of the anomalously fluctuating charge current in the stochastic and deterministic charged cellular automata, which is a classical statistical model first introduced in \cite{SCCA} to describe the dynamics of scattering charged particles on a periodic 1D lattice. 
Ref.\,\cite{YKBI 26} is more relevant to our study, in which the gapped antiferromagnetic XXZ chain is treated with the BMFT adapted so as to capture the spin transport in diffusive scale, which is characterized by the dynamical relationship $\mr{x}=\mr{t}^{1/2}$, and the motion of magnetization concentration is pursued for its initial fluctuation to calculate the variance of the time-integrated spin current which fluctuates at zero magnetic field.

The layout of this paper is as follows. We set up the statistical ensemble which represents the initial state where the spin chain relaxes before removing the magnetic field \eqref{m. field} in Section 2.
After constructing the BMFT equations for the XXZ chain and obtaining the $1/N$-scaled long-range spin correlation function in Section 3, we show section 4 that the relation \eqref{F.D.} holds for the spin conductivity $\s(\beta)$.
Section 5 is devoted to the calculation of $\s(\beta)$ in the limit $\beta\to0$. In particular, we show the divergence of its high temperature proportionality constant at the isotropic point.
Section 6 is a summary and a discussion about the dynamic scaling in spin transport at the isotropic point.
Appendix A recalls the Takahashi-Suzuki (TS) numbers and their associated numbers, which is used to describe the excitations in the critical regime \cite{TS 72} (see also the book \cite{Takahashi 99}).
Appendix B and C contains important details for the BMFT equations and the long-range spin correlation function.


\section{Setup}
We define the Hamiltonian of the spin-1/2 model on a periodic 1D lattice with $N$ sites as
\begin{align}
&H:=J\sum_{k=-N/2+1}^{N/2} \left(S^x_kS^x_{k+1}+S^y_kS^y_{k+1} +\Delta S^z_kS^z_{k+1}\right)
     -(1-\Theta(\mr{t}))U,  \nn\\
&U:=2\sum_{i=1}^{N/L} h_0(\mr{x}_i/N)\sum_{k=0}^{L-1}S^z_{\mr{x}_i-k},
\qquad S^{x,y,z}_{N+1}=S^{x,y,z}_1,
\label{Hamiltonian}\end{align} 
where $S^{x,y,z}_k:=\s^{x,y,z}_k/2$ are Pauli's spin operators at the $k$-th site, $J$ is the coupling constant and $\Delta$ is the anisotropy parameter. The region for the critical regime $0\le \Delta <1$ is parametrized by
\begin{equation}
\Delta=\cos\theta, \quad  \theta=\frac{\pi}{p_0} \quad \mr{and} \quad 2\le p_0<\infty.
\label{anisotropy}\end{equation}
$\Theta(\mr{t})$ denotes the Heaviside step function as
\begin{equation}
\Theta (\mr{t}):=
\begin{dcases}
1 \quad (\mr{t} \ge 0) \\
0 \quad (\mr{t} < 0).
\end{dcases}
\end{equation}
The magnetic field $h_0(\mr{x}/N)$ \eqref{m. field} is turned on at $\mr{t}=-\infty$ and cut off at $\mr{t}=0$.
We introduce a statistical ensemble that represents the initial state of the spin chain under the magnetic field $h_0(\mr{x}/N)$ \eqref{m. field}. The ensemble is specified by the partition function $Z$:
\begin{align}
&Z=\int\left[\prod_i \prod_j \prod_\la 
        d\left(\frac{\varrho_j^{\mr{h};\la}(\mr{x}_i)}{\varrho_j^{(\la)}(\mr{x}_i)}\right)\right]
        \exp\left(-\beta E\,[h_0(\mr{x}_i/N)]
                      +\ms{S}\left[\frac{\varrho_j^{\mr{h};\la}(\mr{x}_i)}{\varrho_j^{(\la)}(\mr{x}_i)}\right]\right),  \label{Z0}\\
&E\,[h(\mr{x}_i/N)]=L\sum_{i=1}^{N/L} e\left[h(\mr{x}_i/N)\right],  \label{E_h} \\
&e\left[h(\mr{x}_i/N)\right]
    =\sum_{j=1}^{m_\alpha} \int_{-\infty}^\infty d\la \left\{\epsilon_j(\la)+2n_jh(\mr{x}_i/N)\right\}\varrho_j^{(\la)}(\mr{x}_i),  \\
&\ms{S}\left[\frac{\varrho_j^{\mr{h};\la}(\mr{x}_i)}{\varrho_j^{(\la)}(\mr{x}_i)}\right]
=L\sum_{i=1}^{N/L}  \ms{s}\left[\frac{\varrho_j^{\mr{h};\la}(\mr{x}_i)}{\varrho_j^{(\la)}(\mr{x}_i)}\right], \label{entropy}\\
&\ms{s}\left[\frac{\varrho_j^{\mr{h};\la}(\mr{x}_i)}{\varrho_j^{(\la)}(\mr{x}_i)}\right]
=
\sum_{j=1}^{m_\alpha}\int_{-\infty}^\infty\! d\la \, \varrho_j^{(\lambda)}(\mr{x}_i)\ln\! \left(1+\frac{\varrho_j^{\mr{h};\lambda}(\mr{x}_i)}{\varrho_j^{(\lambda)}(\mr{x}_i)}\right) + {\varrho_j^{\mr{h};\lambda}(\mr{x}_i)}\ln\! \left(1+\frac{\varrho_j^{(\lambda)}(\mr{x}_i)}{\varrho_j^{\mr{h};\lambda}(\mr{x}_i)}\right)\!.
\end{align}
Here $E[h(\mr{x}_i/N)]$ is the sum of the energies per fluid cell $L e\left[h(\mr{x}_i/N)\right]$, in which $\epsilon_j(\la)$ is the one-particle (string) dispersion that is a function of the momentum $\kappa_j(\la)$ as given in Eq.s \eqref{aj}, $h(\mr{x}_i/N)$ is a magnetic field which creates a profile of the distribution function $\varrho_j^{(\lambda)}(\mr{x}_i)$ of quasiparticles (strings) and $n_j$ is the number of magnons composing $n_j$-string. The set of TS numbers $\{n_j\}_{j=1}^{m_\alpha}$ is uniquely determined by the anisotropy parameter $\theta$ as recalled in Appendix \ref{TS no.}. 
$\ms{S}\left[\varrho_j^{\mr{h};\la}(\mr{x}_i)/\varrho_j^{(\la)}(\mr{x}_i)\right]$ is the sum of the entropies per fluid cell $L \ms{s}\left[\varrho_j^{\mr{h};\la}(\mr{x}_i)/\varrho_j^{(\la)}(\mr{x}_i)\right]$
and $\varrho_j^{\mr{h};\la}(\mr{x}_i)$ is the distribution function of holes of strings.
We rescale the space as
\begin{equation}
x=\mr{x}/N
\label{scaling}\end{equation}
with the length element $dx=L/N$ in the limit \eqref{limits} and rewrite the total energy \eqref{E_h} and entropy \eqref{entropy} as 
\begin{align}
&E[h(\mr{x}_i/N)]=N\int_{-1/2}^{1/2} e\left[h(x)\right]dx,  \nn\\
&\ms{S}\left[\frac{\varrho_j^{\mr{h};\la}(\mr{x}_i)}{\varrho_j^{(\la)}(\mr{x}_i)}\right]
                        = N\int_{-1/2}^{1/2} \ms{s}\left[\frac{\rho_j^{\mr{h};\la}(x)}{\rho_j^{(\la)}(x)}\right] dx, \label{rescaling}\end{align}
where the distribution functions of quasiparticles and holes are rewritten as 
\begin{equation}
\rho_j^{(\lambda)}(x)=\varrho_j^{(\lambda)}(\mr{x}),
\qquad
\rho_j^{\mr{h};\lambda}(x)=\varrho_j^{\mr{h};\lambda}(\mr{x}).
\label{viewpoints}\end{equation}
The spin density is given by
\begin{equation}
s(x)= \frac{1}{2}-m(x), 
\qquad
m(x):=\sum_jn_j\int \rho_j^{(\lambda)}(x) d\la. 
\label{s. density}\end{equation}
Here and henceforth, we may omit the range of sum over the string number and the range of integration over the spectral parameter $\la$ for brevity.
We denote the ensemble average $\bra\bullet\ket_N$ which approximates the observation at the local thermodynamic equilibrium under the field $h_0(x)$. The magnetization density $\bra s(x)\ket_N$ is obtained as 
\begin{align}  
\bra s(x)\ket_N
&=\frac{1}{2}+\frac{1}{2L\beta}\frac{\del \ln Z}{\del h_0(x)}  \nn\\
&=\frac{1}{2}\!-\!\int\!\left[\prod_x \prod_j \prod_\la d\left(\frac{\rho_j^{\mr{h};\la}(x)}{\rho_j^{(\la)}(x)}\right)\!\right]\!\exp\!\left[\!-\beta \left(E[h_0(x)]
                                  -\beta^{-1}\ms{S}\left[\frac{\rho_j^{\mr{h};\la}(x)}{\rho_j^{(\la)}(x)}\right]
                                  -F_0\right)\!\right]\!m(x) \nn\\   
&=\frac{1}{2}-\sum_jn_j\int \rho_j^{(\la)}(x)_0 \,d\la, \nn\\
F_0
&=E[h_0(x)]-\beta^{-1}\ms{S}\left[\frac{\rho_j^{\mr{h};\la}(x)_0}{\rho_j^{(\la)}(x)_0}\right]
\label{min. F}\end{align}
in the limit $N\to\infty$, where the equilibrium distributions $\rho_j^{(\la)}(x)_0$ and $\rho_j^{\mr{h};\la}(x)_0$ are determined by
\begin{equation}
\rho_j^{(\la)}(x)_0=\frac{\varsigma_j}{1+\eta_j^{(\la)}(x)_0}\frac{\del \ln\eta_j^{(\la)}(x)_0}{\del (\beta A)},
   \qquad
\rho_j^{\mr{h};\la}(x)_0=\eta_j^{(\la)}(x)_0 \, \rho_j^{(\la)}(x)_0
\label{rho sol.}\end{equation}
and
\begin{align}
&\varsigma_j:=\sgn\left(a_j(\la)\right),
\quad
a_j(\la):=\frac{\epsilon_j(\la)}{A} =-\frac{\del_\la\kappa_j(\la)}{2\pi}
             =\frac{\theta}{2\pi}\frac{\sin\theta q_j}{\cosh\theta \la + \cos\theta q_j},  \nn\\
&A:=-\frac{2\pi J\sin\theta}{\theta}
\label{aj}\end{align}
with the sequence of numbers $\{q_j\}_{j=1}^{m_\alpha}$ defined in Appendix \ref{TS no.} in relation to the TS numbers.   
In Eq.s \eqref{rho sol.}, $\eta_j^{(\la)}(x)_0$ are the solutions to the thermodynamic Bethe ansatz (TBA) equations \cite{TS 72,Takahashi 99} at $h(x)=h_0(x)$
\begin{align}
&\beta \left\{Aa_j(\la)+2n_jh(x)\right\}
 =\ln\eta_j^{(\la)}(x)-\sum_k\int d\mu\;\varsigma_kT_{jk}(\la-\mu)\ln\left(1+\eta_k^{(\mu)}(x)^{-1}\right);\nn\\ 
&\eta_j^{(\la)}(x):=\frac{\rho_j^{\mr{h};\la}(x)}{\rho_j^{(\la)}(x)},
\label{TBA}\end{align}
where $T_{jk}(\la)$ is a scattering kernel, the explicit expression of which is not used in this paper, though it is given in \cite{TS 72,Takahashi 99}.   
These equations are obtained in the limit $L\to\infty$ from the following saddle point equations for the free energy
$F=E[h(x)]-\beta^{-1}\ms{S}\left[\rho_j^{\mr{h};\la}(x)/\rho_j^{(\la)}(x)\right]$: 
\begin{align}
\frac{1}{L}\frac{\delta\, F}{\delta \rho_j^{(\la)}(x)}
=
Aa_j(\la)+2n_jh(x)-\frac{1}{L\beta}\frac{\delta}{\delta \rho_j^{(\la)}(x)}
\ms{S}\left[\frac{\rho_j^{\mr{h};\la}(x)}{\rho_j^{(\la)}(x)}\right]
=0.
\label{saddle 0}\end{align}
At the applied magnetic field ($h(x)=h_0(x)$), the value of $F$ at the saddle point is the $F_0$ in \eqref{min. F}.

We define the spin correlation function by
\begin{equation}
\bra s(x)s(x') \ket^\ms{c}_N:=\bra s(x)s(x') \ket_N-\bra s(x) \ket\bra s(x') \ket_N.
\label{ccf}\end{equation}
In the initial state, the long-range spin correlation is absent as
\begin{align}
\bra s(x)s(x')\ket^\ms{c}_N
&=
\frac{1}{2L\beta}\frac{\del \bra s(x) \ket_N}{\del h_0(x')}  \nn\\
&=
-\frac{1}{2L\beta}\sum_jn_j\int \frac{\del \rho_j^{(\la)}(x)_0}{\del h_0(x')}d\la
=0 
\quad \textup{except for} \;\; x=x'.
\label{initial correlation}\end{align}

\section{BMFT equations and long-range spin correlation}
In the following, we construct BMFT equations for the present system and show that the long-range spin correlation develops after cutting off the field $h_0(x)$ at $\mr{t}=0$.
\subsection{BMFT equations}
At first, we employ the formulation of generalized hydrodynamics (GHD) that the equations of motion \eqref{eq.o.m} take the form of the Euler equations for quasiparticle densities $\rho_j^{(\la)}(x,t)$ 
as \cite{ND 16,Doyon 16,BVKM}  
\begin{align} 
\del_t \rho_j^{(\la)}(x,t)+\del_x \left(v_j^{\mathrm{eff};\la}(x,t)\rho_j^{(\la)}(x,t)\right)=0
\label{eq. of motion}\end{align}
with the initial values $\rho_j^{(\la)}(x,0)=\rho_j^{(\la)}(x)$.
In these equations, the quasiparticle current appears in ballistic scale, which is in order $O(1)$, and its $1/N$-scaled diffusive current is neglected.
The solutions to these equations compose the motion of spin density as
\begin{equation}
s(x,t)= \frac{1}{2}-m(x,t), 
\qquad
m(x,t):=\sum_jn_j\int \rho_j^{(\lambda)}(x,t) d\la. 
\end{equation}
Here, time $\mr{t}$ is rescaled corresponding to the spatial rescaling \eqref{scaling} as 
\begin{equation}
t=\mr{t}/N
\label{t scale}\end{equation}
and $v^\mathrm{eff;\la}_j(x,t)$ is the motion of the effective velocity,
\begin{equation}
v^\mathrm{eff;\la}_j(x,t)
:=\frac{\sum_k(R^\ms{-T}_{jk})^\la_{\;\,\mu}(x,t) \,\del_\mu\epsilon_k^{(\mu)}}
          {\sum_k(R^\ms{-T}_{jk})^\la_{\;\,\mu}(x,t) \,\del_\mu\kappa_k^{(\mu)}}
\label{eff. v}\end{equation}
where $(R^\ms{-T}_{jk})^\la_{\;\,\mu}(x,t)$ is the kernel of the inverse of integral operator $R^\ms{T}(x,t)$ which is defined in the dressing equations \cite{Korepin}.
It is necessary that the motion of particle and hole densities of strings satisfy the following dressing equations: \begin{align}
&
a_j(\la)
=\varsigma_j \{\rho_j^{(\la)}(x,t)+\rho_j^{\mr{h};\la}(x,t)\}+ \sum_k\int d\mu\,T_{jk}(\la-\mu) \rho_k^{(\mu)}(x,t)  \nn\\
&
\hspace{0.9cm}=:\sum_k(R^\ms{T}_{jk})^\la_{\;\,\mu}(x,t) \varsigma_k\rho_k^{\mr{t};\mu}(x,t),  \label{rho's dress}\\
&
\quad   R(x,t)= 1+\varsigma\vartheta(x,t) T,   
\qquad R^\ms{T}(x,t)= 1+ T \varsigma \vartheta(x,t)  \label{rhot}
\end{align}
and
\begin{equation}
\rho_j^{\ms{t};\la}(x,t):=\rho_j^{(\la)}(x,t)+\rho_j^{\mr{h};\la}(x,t),    
\qquad
\vartheta_j^{(\la)}(x,t):=\frac{\rho_j^{(\la)}(x,t)}{\rho_j^{\mr{t};\la}(x,t)}.
\label{Fermi w.}\end{equation}
In the following, $f_j^{\dr;\la}(x,t):=\sum_k(R^\ms{-T}_{jk})^\la_{\;\,\mu}(x,t)f_k^{(\mu)}$ universally means a dressed charge of arbitrary bare charge $f_j(\la)$ for the $j$-th quasiparticles. 
The equations \eqref{rho's dress} are fundamental in the present system because these equations lead to the equations of motion of density ratios $\eta_j^{(\la)}(x,t)=\rho_j^{\mr{h};\la}(x,t)/\rho_j^{(\la)}(x,t)$,
\begin{equation}
\del_t \eta_j^{(\la)}(x,t)+ v_j^{\mr{eff};\la}(x,t) \del_x  \eta_j^{(\la)}(x,t)=0
\label{eta law}\end{equation}
with the initial values $\eta_j^{(\la)}(x,0)=\eta_j^{(\la)}(x)$ determined by the TBA equations \eqref{TBA}.
The proof of Eq.s \,\eqref{eta law} is as follows:
\begin{proof}
The Euler-scaled GHD equations of motion \eqref{eq. of motion} are transformed as
\begin{align}
0
&=\del_t \left[\vartheta_j^{(\la)}\rho_j^{\mr{t};\la}\right](x,t)
    -\del_x \left[\vartheta_j^{(\la)}\frac{\del\epsilon_j^{\dr;\la}}{2\pi\varsigma_j}\right](x,t)    \nn\\
&=\sum_k\left[(R_{j,k}^{-1})^\la_{\;\,\mu} \rho_k^{\ms{t};\mu}\del_t\vartheta_k^{(\mu)}\right](x,t)
    -\sum_k\left[(R_{j,k}^{-1})^\la_{\;\,\mu}\frac{\del\epsilon_k^{\dr;\mu}}{2\pi\varsigma_k}\del_x \vartheta_k^{(\mu)}\right](x,t) \nn\\
&=\sum_k\left[(R_{j,k}^{-1})^\la_{\;\mu} \rho_k^{\ms{t};\mu}\right](x,t)\left(\del_t\vartheta_k^{(\mu)}(x,t)+ v_k^{\mr{eff};\mu}(x,t) \del_x \vartheta_k^{(\mu)}(x,t)\right),
\label{transform}\end{align}
where the symbol $[\cdots](x,t)$ denotes that the factors in the square bracket are functions of $x, t$. 
In the first line of Eq.\,\eqref{transform}, we use the definition of the motion of the Fermi weight $\vartheta^{(\la)}_j(x,t)$ \eqref{Fermi w.}, and we have introduced a notation $\del f=\del_\la f_j(\la)$ to rewrite Eq.\,\eqref{eff. v} as
\begin{equation}
v_j^{\mr{eff};\la}(x,t)=-\left[\frac{\del\epsilon_j^{\dr;\la}}{2\pi\varsigma_j\rho_j^{\ms{t};\la}}\right]\!(x,t)
\label{ev}\end{equation}
by combining the relations $\del_\la\kappa_j(\la)=-2\pi a_j(\la)$ from Eq\,\eqref{aj} and
$a_j^{\dr;\la}(x,t)=\varsigma_j\rho_j^{\ms{t};\la}(x,t)$ from Eq.\,\eqref{rho's dress}.
To obtain the second line of Eq.\,\eqref{transform}, we define an operator  
$\mathfrak{U}(x,t):=[\vartheta R^{-\ms{T}}](x,t)$ and use a dressed function formula  
\begin{align}
\del_p \left[\vartheta_j^{(\la)}f_j^{\dr;\la}\right]\!(x,t)
&=\sum_k\del_p(\mathfrak{U}_{jk})^\la_{\;\,\mu}(x,t)f_k^{(\mu)}  \nn\\
&=\varsigma_j\sum_{k,l} \left[(R^{-1}_{jk})^\la_{\;\,\mu}\varsigma_k\left\{\del_p\vartheta_k^{(\mu)}\right\}(R^{-\ms{T}}_{kl})^\mu_{\;\,\nu}\right](x,t)f_l^{(\nu)}  \nn\\
&=\varsigma_j\sum_k \left[(R^{-1}_{jk})^\la_{\;\,\mu}\varsigma_k\left\{\del_p\vartheta_k^{(\mu)}\right\}f_k^{\dr;\mu}\right](x,t).
\label{dff}\end{align}
where $p$ is a parameter or variable which is chosen as $p=\beta, \beta A, 2\beta h, x$ or $t$ and $p\ne\la$.
This formula was given in \cite{Doyon 16} (see Eq.s (24) and (33) in this reference) but now generalized for the present system which contains multiple species of quasiparticles.
From the last expression of Eq.\,\,\eqref{transform}, we obtain the equations of motion \eqref{eta law} using the chain rule
$\del_{x,t}\eta_j^{(\la)}(x,t)=[d\eta_j^{(\la)}/d\vartheta_j^{(\la)}](x,t)\del_{x,t}\vartheta_j^{(\la)}(x,t)$.
\end{proof}
The GHD equations of motion \eqref{eq. of motion} and \eqref{eta law} are free from any force term \cite{force} in the time evolution without external field, which makes $\eta^{(\la)}_j(x,t)$ and arbitrary dressed charge $f^{\dr;\la}_j(x,t)$ constant during motion in the substantial viewpoint of hydrodynamics at velocity $dx/dt=v^{\mr{eff};\la}_j(x,t)$ as
$d f_j^{\dr;\la}(x,t)/dt=[df_j^{\dr;\la}/d\eta_j^{(\la)}](x,t) (\del_t \eta_j^{(\la)}(x,t)+ v_j^{\mr{eff};\la} \del_x \eta_j^{(\la)}(x,t))=0$.
Therefore, the solutions to the equations of motion \eqref{eta law} can be expressed by the initial values $\eta_j^{(\la)}(x)$ and dressed charges are determined as
\begin{equation}
\eta_j^{(\la)}(x,t)=\eta_j^{(\la)}\!\left(u^{(\la)}_j(x,t)\right),
\qquad
f^{\dr;\la}_j(x,t)=f_j^{\dr;\la}\!\left(u^{(\la)}_j(x,t)\right),
\label{eta sol.}\end{equation} 
where $u_j^{(\la)}(x,t)$ points the local space from which the physical excitation, that is described by the dressed charges of $n_j$-string with spectral parameter $\la$---namely $f^{\dr;\la}_j(x,t)$, starts the substantial motion toward a space-time point $(x,t)$.
The initial position $u^{(\la)}_j(x,t)$ is determined by the following integral equation \cite{DSY curve} (see also \cite{Doyon lec.}):
\begin{equation}
\int_{-1/2}^x\rho_j^{\ms{t};\la}(y,t)\,dy - \int_{-1/2}^{u_j^{(\la)}(x,t)}\rho_j^{\ms{t};\la}(y,0)\,dy 
= \hat{v}_j(\la)\,t
\label{characteristics}\end{equation}
with $\hat{v}_j(\la):=\lim_{x\to-1/2}v_j^{\mr{eff};\la}(x,0)\rho_j^{\ms{t};\la}(x,0)$.   
Using the first identity in Eq.s\,\eqref{eta sol.}, one can rewrite Eq.\,\eqref{eff. v} as
\begin{equation}
v^\mathrm{eff;\la}_j(x,t)=-\frac{A}{2\pi}\frac{\del}{\del_\la}\ln\!\left(\del_{\beta A}\ln\eta_j^{(\la)}\!\left(u^{(\la)}_j(x,t)\right)\right),
\label{eos}\end{equation}
which is seen from Eq.\,\eqref{aj} $\del_\la\epsilon_j(\la)=A\del_\la a_j(\la)$ and differentiating the TBA equations \eqref{TBA} with respect to $\beta A$ to have
$a_j^{\dr;\la}(x)=\del_{\beta A}\ln\eta_j^{(\la)}(x)$. 
This is an equation of state which states that the effective velocity is fully determined by the density ratios of strings.

Following the central concept of BMFT, 
we approximate the observation for the evolution process by means of the Euler-scaled GHD equations \eqref{eq. of motion}. The approximated value $\bbra\bullet\kket_N$ is given by
\begin{align}
&\bbra \bullet \kket_{N}
 =Z^{-1}\int \left[\prod_{x,t,j,\la} d\left(\frac{\rho_j^{\mr{h};\la}(x,t)}{\rho_j^{(\la)}(x,t)}\right)\right]
    e^{-\beta F[h_0(x)]}
    \int\left[\prod_{x,t} g_j^{(\la)}(x,t)\right]e^{-\beta X_1 }\bullet, \nn\\
&F[h_0(x)]=E[h_0(x)]-\beta^{-1}\ms{S}\left[\frac{\rho_j^{\mr{h};\la}(x)}{\rho_j^{(\la)}(x)}\right],
\nn\\
&X_1=2N \sum\limits_jn_j\int_{-1/2}^{1/2} dx\int_0^\tau \!dt \int \!d\la\, 
     g_j^{(\la)}(x,t)
     \left\{\del_t \rho_j^{(\lambda)}(x,t)
     +\del_x \left(v_j^{\mathrm{eff};\lambda}(x,t)\rho_j^{(\lambda)}(x,t) \!\right) \! \right\},
\label{relaxing s}\end{align}
where $\tau$ is the rescaled time which has been introduced in Eq.\,\eqref{F.D.} and $g_j^{(\la)}(x,t)$ are the auxiliary fields which imposes Euler's equations \eqref{eq. of motion} on the state evolution.
We calculate the motion of the spin correlation function \eqref{ccf} as 
\begin{align}
\bbra s(x,t)s(x',t')\kket^\ms{c}_N
&=\frac{1}{(2\beta)^2}
    \left.
    \left(\frac{\del^2 \bbra e^{2\beta  X_2} \kket_N}{\del c\del c'}
    -\frac{\del \bbra e^{2\beta  X_2} \kket_N}{\del c}\frac{\del \bbra e^{2\beta  X_2} \kket_N}{\del c'}\right)
    \right|_{c=c'=0}  \nn\\
&=\frac{1}{(2\beta )^2}\left.\frac{\del}{\del c'}\left(\left.\frac{\del}{\del c}\ln\bbra e^{2\beta X_2} 
    \kket_N\right|_{c=0}\right)\right|_{c'=0}  \nn\\
&=-\frac{1}{2\beta }\sum_j n_j\int \left.\frac{\del}{\del c'}
     \bbra\rho_j^{(\la)}(x,t)\kket_{N,c'}\right|_{c'=0} d\la,   
\label{cscf}\end{align}
where $X_2=c s(x,t)+c' s(x',t')$ with arbitrary constants $c, c'$,
and $\bbra\rho_j^{(\la)}(x,t)\kket_{N,c'}$ is a function of $c'$ and determined in the limit $N\to\infty$ by the BMFT equations \cite{BMFT 1, BMFT 2} to which the equations of motion \eqref{eta law} and the followings belong in the present system: 
\begin{align}
&h_0(x)-h(x)=g_j^{(\la)}(x,0) =:g(x),  \label{saddle eq.1}\\
&g_j^{(\la)}(\pm1/2,t)=0, \qquad g_j^{(\la)}(x,\tau)=0,  \label{saddle eq.2}\\
&\del_{t} g_j^{(\la)}(x,t)+\left[v_j^{\mr{eff};\la}\del_{x} g_j^{(\la)}\right]\!(x,t)=N^{-1}c' \delta(x-x')\delta(t-t'). \label{saddle eq.3}\end{align}
Eq.s \eqref{saddle eq.1}-\eqref{saddle eq.3} are obtained from the saddle point equations
\begin{equation}
\frac{\delta}{\delta \rho_j^{(\lambda)}(x,t)}\Big(F[h_0(x)]+X_1-2  c' s(x',t')\Big)=0. 
\label{saddle}\end{equation}
To obtain the initial condition \eqref{saddle eq.1}, we use Eq.s \eqref{saddle 0} in conjunction with \eqref{saddle}. 
We show Appendix \ref{F. Jacobian} that Eq.\,\eqref{saddle eq.3} is derived from Eq.\,\eqref{saddle} by using the flux Jacobian $\mathcal{J}(x,t)$ with the kernel
\begin{equation}
(\mathcal{J}_{jk})^\la_{\;\mu}(x,t)
  :=\frac{\delta\left[v_j^{\mr{eff};\la}\rho_j^{(\la)}\right]\!(x,t)}{\delta \rho_k^{(\mu)}(x,t)}, 
\label{Jacobian}\end{equation}
which is diagonalized as
\begin{equation}
\sum_{j,k}\left[(R_{lj})^\nu_{\;\,\la}(\mathcal{J}_{jk})^\la_{\;\,\mu}   (R^{-1}_{kl})^\mu_{\;\,\nu}\right]\!(x,t)
=v_l^{\mr{eff};\nu}(x,t).
\label{d.Jaco}\end{equation}

\subsection{Long-range spin correlation}
The spin correlation function \eqref{cscf} is calculated as
\begin{align}
\bbra s(x,t)s(x',t')\kket^\ms{c}_N
&=-\frac{1}{2\beta }\sum_j n_j\int d\la\left.\left[(R_{jk}^{-1})^\la_{\;\,\mu}\rho_k^{\ms{t};\mu}\right]
      \!(x,t)_0 \, \del_{c'}\vartheta_k^{(\mu)}(x,t)\right|_{c'=0}  \nn\\
&=\frac{1}{2\beta}\sum_j \int d\la
    \left.\left[n_j^{\dr;\la} \Gamma_j^{(\la)}\right]\!(x,t)_0
    \del_{c'} \ln\eta_j^{(\la)}(x,t)\right|_{c'=0}. 
\label{cscf 2}\end{align}
and
\begin{equation}
\Gamma_j^{(\la)}(x,t):= \varsigma_j\left[\frac{\del_{\beta A} \ln \eta_j^{(\la)}}
                                {\left(1+\eta_j^{(\la)}\right)\left(1+\left(\eta_j^{(\la)}\right)^{-1}\right)} \right]\!(x,t).
\label{density}\end{equation}
Here and henceforth any $c'$-dependent function $a(x,t)$ and its initial values $a(x)=a(x,0)$ are denoted by $a(x,t)_0$ and $a(x)_0$ at $c'=0$, or equivalently at $h(x)=h_0(x)$ as seen from Eq.s \eqref{saddle eq.1} and \eqref{saddle eq.3}.
For the second expression of Eq.\,\eqref{cscf 2}, we use the transformations which were performed in Eq.\,\eqref{transform}.
For Eq.\,\eqref{density} we use the identity
$\rho_j^{\ms{t};\la}(x,t)=\varsigma_j\del_{\beta A}\ln\eta_j^{(\la)}(x,t)$.
In the last expression of Eq.\,\eqref{cscf 2}, $n_j^{\dr;\la}(x,t)$ is the one-particle magnetization which is given by
\begin{equation}
n_j^{\dr;\la}(x,t)=n_j^{\dr;\la}\!\left(u_j^{(\la)}(x,t)\right)
                     =\frac{\del\ln\eta_j^{(\la)}\!\left(u_j^{(\la)}(x,t)\right)}{\del 2\beta h\!\left(u_j^{(\la)}(x,t)\right)},
\label{1 mag}\end{equation}
and the dressed charge $\del_{c'}\ln\eta_j^{(\la)}(u_j^{(\la)}(x,t))$ is rewritten as
\begin{align}
\del_{c'}\ln\eta_j^{(\la)}(x,t)
&=\del_{c'}\ln\eta_j^{(\la)}\!\left(u_j^{(\la)}(x,t)\right)  \nn\\
&=2\beta\sum_k(R_{jk}^{-\ms{T}})^\la_{\;\,\mu}\!\left(u_k^{(\mu)}(x,t)\right) n_k 
   \del_{c'}\!\left\{h_0\left(u^{(\mu)}_k(x,t)\right) -g\!\left(u^{(\mu)}_k(x,t)\right)\right\}  \nn\\
&=\frac{\del \ln\eta_j^{(\la)}\!\left(u_j^{(\la)}(x,t)\right)}{\del u_j^{(\la)}(x,t)}\del_{c'}u_j^{(\la)}(x,t) \nn\\      
&   \quad -2\beta\sum_k n_k \!\left[(R_{jk}^{-\ms{T}})^\la_{\;\,\mu} \del_{c'}g \right]\! 
                \left(u^{(\mu)}_k(x,t)\right).  
\label{3rd factor}\end{align}
The second expression of Eq.\,\eqref{3rd factor} is derived from Eq.\,\eqref{eta sol.} and subsequent expressions are seen from differentiating the TBA equations \eqref{TBA} with respect to $c'$ and using the initial condition \eqref{saddle eq.1}. We showed Appendix \ref{App.D} that this dressed charge is obtained as follows at $c'=0$ and $t=t'$, addressing the equal-time spin correlation: 
\begin{align}
\left.\del_{c'}\ln\eta_j^{(\la)}(x,t')\right|_{c'=0}  
&=
2\beta N^{-1}\left\{n_j^{\dr;\la}({u'}^\la_j)_0\,\delta(x-x') 
+ \mathfrak{S}_j^{(\la)}(x,t')_0  \right\}.  
\label{version}\end{align}
Here and henceforth we use the notations ${u'}_j^{\,\la}:=u_j^{(\la)}(x',t')_0$ and ${u}_j^{\,\la}:=u_j^{(\la)}(x,t)_0$.
The $\mathfrak{S}_j^{(\la)}(x,t')_0$ is determined by the linear integral equations
\begin{equation}
\mathfrak{S}_j^{(\la)}(x,t')_0
=\mathfrak{S}_{0;j}^{(\la)}(x,t')_0  +\varsigma_j\mathcal{W}_j^{(\la)}\!\left(u_j^{(\la)}(x,t')\right)_0 \sum_k
  \int_{-1/2}^x \!dy\, \!\left[(T^\dr_{jk})^\la_{\;\,\mu} \Gamma_k^{(\mu)}
   \mathfrak{S}_k^{(\mu)}\right]\!(y,t')_0 
\label{version 1}\end{equation}
with driving terms
\begin{align}
&\mathfrak{S}_{0;j}^{(\la)}(x,t')_0=d_j^{(\la)}(x,t')_0+{d'}_j^{(\la)}(x,t')_0,  \nn\\
&d_j^{(\la)}(x,t')_0
:=
\varsigma_j\mathcal{W}_j^{(\la)}\!\left(u_j^{(\la)}(x,t')_0\right)_0\sum_k 
\left[(T^\dr_{jk})^\la_{\;\,\mu}\Gamma_k^{(\mu)} n_k^{\dr; \mu}\right]\!({u'}_k^\mu)_0\,\Theta(x-x'), \nn\\
&{d'}_j^{(\la)}(x,t')_0
:=
-\varsigma_j\mathcal{W}_j^{(\la)}\!\left(u_j^{(\la)}(x,t')_0\right)_0\sum_k 
\left[(T^\dr_{jk})^\la_{\;\,\mu} \Gamma_k^{(\mu)} n_k^{\dr;\mu}\right]\!(x')_0\,\Theta\!\left(u_j^{(\la)}(x,t')_0-x'\right),
\label{d. term}\end{align}
where
\begin{equation}
\mathcal{W}_j^{(\la)}(x)_0
:=\left[\frac{\del_x \ln\eta_j^{(\la)}}{\rho_j^{\ms{t};\la}}\right]\!(x)_0. 
\label{version 2}\end{equation}
We formulated the equations \eqref{version 1} for the present system corresponding to Eq.\.(133) in Ref.\,\cite{BMFT 1}, which is a part of the main results from the BMFT for the correlation function.   
Inserting Eq.\,\eqref{version} into the spin correlation function \eqref{cscf 2}, we have 
\begin{equation}
\bbra s(x,t')s(x',t')\kket^\ms{c}_N=N^{-1}\left\{\beta^{-1}\chi(x,t';\beta)\delta(x-x')+S(x,x';t')\right\}, \label{version end}\end{equation}
where
\begin{align}
\chi(x,t;\beta)
&:=N\int_{-1/2}^{1/2}dx'\bbra s(x,t')s(x',t')\kket^\ms{c}_N -\int_{-1/2}^{1/2}dx'S(x,x';t')  \nn\\
&=\beta\sum_j \int d\la\left[(n_j^{\dr;\la})^2\Gamma_j^{(\la)}\right]\!(x,t)_0
\label{s.m.sus}\end{align}
is the static magnetic susceptibility per site \cite{Eggert}
and
\begin{equation}
S(x,x';t')/N:=\sum_j \int d\la
             \left[n_j^{\dr;\la}\Gamma_j^{(\la)}\mathfrak{S}_j^{(\la)}\right](x,t')_0/N, 
\label{lrscf}\end{equation}
is the $1/N$-scaled long-range spin correlation function.

\section{Relation for $\s(\beta)$}\label{Sec.s}
We have introduced the spin conductivity $\s(\beta)$ and proposed its relation with the long-range spin correlation in Eq.\,\eqref{F.D.}. 
In the following, we first define the operators of the spin density $\mr{q}_0(\mr{x})$ and its current $\mr{j}_0(\mr{x})$ and translate the operator description of $\s(\beta)$ into the quasiparticle description. We then show that $\s(\beta)$ is the integration of the $1/N$-scaled long-range spin correlation function.

Let us define the above density operators by  
\begin{align}
&\mr{q}_0(\mr{x}):=L^{-1}\sum_{k=\mr{x}-L+1}^\mr{x}S^z_k,  \nn\\ 
&\mr{j}_0(\mr{x}):=L^{-1} \sum_{k=\mr{x}-L+1}^\mr{x} \mathscr{J}_k,  \qquad
\mathscr{J}_k:=i\frac{J}{2} (S^+_k S^-_{k+1}-S^-_kS^+_{k+1}) \,; \quad S^\pm_k=S^x_k \pm iS^y_k.
\end{align}
Total spin $S=L\sum_{i=1}^{N/L}\mr{q}_0(\mr{x}_i)$ is conserved as $[S,H]=0$ with the Hamiltonian $H$ defined by \eqref{Hamiltonian}.
Let $\mr{o}(\mr{x},\mr{t})$ be the Heisenberg representation of arbitrary density operator $\mr{o}(\mr{x})$ following the equation
\begin{equation}
\frac{\del\mr{o}(\mr{x},\mr{t})}{\del \mr{t}}=\frac{1}{i}[\mr{o}(\mr{x},\mr{t}), H],
 \quad
 \mr{o}(\mr{x},0)=\mr{o}(\mr{x}).
\label{Heisenberg}\end{equation}
We define the differentiation of $\mr{o}(\mr{x},\mr{t})$ with respect to $\mr{x}$ by
\begin{equation}
\frac{\del \mr{o}(\mr{x},\mr{t})}{\del\mr{x}}:=\mr{o}(\mr{x},\mr{t})-\mr{o}(\mr{x}-1,\mr{t}).
\label{derivative}\end{equation}
For then, using the discrete continuity equation $-i[S_k^z,H]+(\mathscr{J}_k-\mathscr{J}_{k-1})=0$, we obtain the hydrodynamic equation of motion of the spin as
\begin{equation}
\del_\mr{t}\mr{q}_0(\mr{x},\mr{t})+\del_\mr{x} \mr{j}_0(\mr{x},\mr{t})=0.
\label{hydro. eq}\end{equation}
We rescale the space and time as $x=\mr{x}/N$ \eqref{scaling} and $t=\mr{t}/N$ \eqref{t scale} respectively to rewrite the density operator as 
\begin{equation}
o(x,t) = \mr{o}(\mr{x},\mr{t}).
\label{notation change}\end{equation} 
We also rescale the space and time differentiation \eqref{derivative} and \eqref{Heisenberg} as
\begin{equation}
\frac{\del o(x,t)}{\del x}=\frac{N}{L} \frac{\del \mr{o}(\mr{x},\mr{t})}{\del\mr{x}},
\qquad
\frac{\del o(x,t)}{\del t}=\frac{N}{L}  \frac{\del \mr{o}(\mr{x},\mr{t})}{\del\mr{t}}
\label{difference}\end{equation}
and rewrite the equation of motion \eqref{hydro. eq} as
\begin{equation}
\del_t q_0(x,t)+\del_x j_0(x,t)=0.
\label{hydro. eq2}\end{equation}
The spin current density is expressed as
\begin{equation}
j_0(x,t)=\mr{j}_0(\mr{x},\mr{t})=\frac{\del\mathscr{S}(\mr{x},\mr{t})}{\del\mr{t}} 
=
L\left(x-\frac{L}{2N}\right)\frac{\del q_0(x,t)}{\del t},
\label{q-j}\end{equation}
where $\mathscr{S}(\mr{x})\!:=L^{-1}\sum_{k=\mr{x}-L+1}^\mr{x} k S_k^z$.
The last expression of Eq.\,\eqref{q-j} is valid in the limit $L/N\!\to0$.

Using the space-time rescaling \eqref{scaling} and \eqref{t scale}, the spin conductivity \eqref{F.D.} is rewritten as
\begin{equation}
\s(\beta)=\frac{\beta N^2}{L^2\tau}\int_0^\tau dt \int_0^\tau dt' 
               \int_{-1/2}^{1/2}dx\bbra j_0(0,t')j_0(x,t)\kket^\ms{c}.
\label{ass.}\end{equation}
where the superscript $^\ms{c}$ denotes the connected correlation function as defined by Eq. \eqref{ccf}.
At the thermodynamic equilibrium, $\s(\beta)$ is the spin dc conductivity based on Kubo's linear response theory \cite{Kubo} (see also Appendix A in \cite{Ae 2024}) as
\begin{align}
\s(\beta)
&=\frac{\beta N^2}{2L}\int_{-\tau}^\tau dt \int_{-1/2}^{1/2}dx\bra j_0(0,0)j_0(x,t)\ket  \nn\\
&=\lim_{\mathcal{T}\to\infty} D_s(\beta)\mathcal{T}+\s^\mr{reg}(\beta) 
\qquad \textup{at the thermodynamic equilibrium}, 
\label{linear resp}\end{align}
where $\bra\cdots\ket$ denotes the ensemble average which approximates the observation for the system in this equilibrium. The spin flux is not observed in this state---namely, $\bra j_0(x,t)\ket=0$.
$D_s(\beta)$ is the finite temperature spin Drude weight, which is nonzero in the critical regime \eqref{anisotropy} \cite{Zotos,Klumper,AS}.
$\s^\mr{reg}(\beta)$ is the regular part of the spin dc conductivity.
On the other hand, Eq.\,\eqref{ass.} is transformed as follows in the non-equilibrium state after removing the magnetic field $h_0(x)$ \eqref{m. field}:    
\begin{align}   
\s(\beta)
&=-\frac{\beta N^2}{L^2\tau}\int_0^\tau dt \int_0^\tau dt' \int_{-1/2}^{1/2} dx 
   x \bbra j_0(0,t') \del_x j_0(x,t)\kket^\ms{c}  \nn\\
&=-\frac{\beta N}{2\tau}\int_0^\tau dt
    \int_0^\tau dt' \int_{-1/2}^{1/2} dx 
   x \bbra \del_tq_0(0,t') \del_t q_0(x,t)\kket^\ms{c}  \nn\\
&=-\frac{\beta N}{2\tau}\int_{-1/2}^{1/2} dx 
   x \bbra \{q_0(0,\tau)-q_0(0,0) \}\{ q_0(x,\tau) -q_0(x,0) \}\kket^\ms{c}  \nn\\
&=-\frac{\beta N}{2\tau}\int_{-1/2}^{1/2} dx 
   x \biggl(\bbra  q_0(0,\tau)q_0(x,\tau) \kket^\ms{c} +\bbra q_0(0,0)q_0(x,0) \kket^\ms{c} \nn\\
&\quad  -\bbra  q_0(0,0)q_0(x,\tau) \kket^\ms{c} -\bbra  q_0(0,\tau)q_0(x,0) \kket^\ms{c}\biggr).         
\label{j to q 0}\end{align}
In the second expression of Eq. \eqref{j to q 0}, the integration by parts is used.
In the third expression of the same, Eq.s \eqref{hydro. eq2} and \eqref{q-j} are used.
Using $\bbra s(x,t)s(x,t')\kket^\ms{c}_N$, which is the quasiparticle picture of the spin correlation function $\bbra  q_0(x',t')q_0(x,t) \kket^\ms{c}$  , $\s(\beta)$ is obtained as
\begin{align}
\s(\beta)
&=-\frac{N\beta}{2\tau}\int_{-1/2}^{1/2}dxx
   \Bigl((\beta N)^{-1}\chi(x,\tau;\beta)\delta(x)  +S(0,x;\tau)/N +\bbra s(0,0)s(x,0)\kket^\ms{c}_N  \nn\\
&\quad -\bbra s(0,0)s(x,\tau)\kket^\ms{c}_N-\bbra s(x,0)s(0,\tau)\kket^\ms{c}_N\Bigr)  \nn\\
&=-\frac{\beta}{2\tau}\int_{-1/2}^{1/2}dxx S(0,x;\tau),
\label{sigma}\end{align}
where the static magnetic susceptibility $\chi(x,t;\beta)$ and the $1/N$-scaled long range spin correlation function $S(x,x';t')$ are given by \eqref{s.m.sus} and \eqref{lrscf} respectively.
The rest of this section is devoted to the proof of Eq.\,\eqref{sigma}:

We start from integrating the both sides of Eq.\,\eqref{saddle eq.3} over $t$'s interval $[t'-\epsilon,t'+\epsilon]$ with $\epsilon$ infinitesimal, which yields
\begin{equation}
g_j^{(\la)}(x,t'+\varepsilon)-g_j^{(\la)}(x,t'-\varepsilon)=N^{-1}c' \delta(x-x').
\label{g's eq0}\end{equation}
From the condition $g_j^{(\la)}(x,\tau)=0$ \eqref{saddle eq.2} and Eq.\,\eqref{saddle eq.3} which makes $g_j^{(\la)}(x,t)$ constant in the motion at $dx/dt=v_j^{\mr{eff};\la}(x,t)$ for $t\not\in [t'-\epsilon,t'+\epsilon]$, we have
\begin{equation}
g_j^{(\la)}(x,t'-\epsilon)=g\!\left(u_j^{(\la)}(x,t'-\epsilon)\right),
\qquad
g_j^{(\la)}(x,t'+\epsilon)=0,
\label{g0}\end{equation}
Using the second identity in Eq.\,\eqref{g0} and multiplying the both sides of Eq.\,\eqref{g's eq0} by $n_j$, we have
\begin{equation}
n_jg_j^{(\la)}(x,t_\epsilon)=-N^{-1}c'n_j\delta(x-x').
\label{g's sol}\end{equation}
where $t_\epsilon:=t'-\epsilon$. From the first identity in Eq.\,\eqref{g0}, this is rewritten as
\begin{align}
n_j g^{(\la)}_j \!\left(\mathcal{U}_j^{(\la)}(x,0;t_\epsilon),t_\epsilon\right)
         &=-N^{-1}c'n_j\delta\!\left(\mathcal{U}_j^{(\la)}(x,0;t_\epsilon)-x'\right),
\label{njg}\end{align}
where $\mathcal{U}_j^{(\la)}(x,0;t)$ is the position satisfying the equation $\eta_j^{\la}\!\left(\mathcal{U}_j^{(\la)}(x,0;t),t\right)=\eta_j^{\la}(x)$ \eqref{eta sol.}.
Dressing the derivatives with respect to $c'$ of the both sides of Eq.\,\eqref{njg} and using the identity for dressed charges $f^{\dr;\la}_j(x,t)=f_j^{\dr;\la}(u^{(\la)}_j(x,t))$ \eqref{eta sol.}, we have
\begin{align}
\sum_k n_k \!\left[(R^{-\ms{T}}_{jk})^\la_{\;\,\mu} \del_{c'} g\right]\!(x)
&=-N^{-1} \sum_k  (R^{-\ms{T}}_{jk})^\la_{\;\,\mu}({u'}_k^\mu)n_k \delta
   \!\left(\mathcal{U}_k^{(\mu)}(x,0;t')-x'\right)
\label{additional}\end{align}
in the limit $\epsilon\to0$.
Because $\del_{c'}u_j^{(\la)}(x,0)=\del_{c'}x=0$, inserting Eq\,\eqref{additional} into the dressed charge \eqref{3rd factor} yields
\begin{align}
\del_{c'}\ln\eta_j^{(\la)}(x)  
&=2\beta N^{-1} \sum_k  (R^{-\ms{T}}_{jk})^\la_{\;\,\mu}({u'}_k^{\mu})n_k 
   \delta\!\left(\mathcal{U}_k^{(\mu)}(x,0;t')-x'\right).
\end{align}
Inserting this into the spin correlation function \eqref{cscf 2}, we obtain 
\begin{align}
\bbra s(x,0)s(x',t')\kket^\ms{c}_N
&=N^{-1}\sum_{j,k} \int d\la \left[n_j^{\dr;\la} \Gamma_j^{(\la)}\right] (x)_0   \nn\\
&\qquad\qquad\qquad\qquad\times
   (R^{-\ms{T}}_{jk})^\la_{\;\,\mu}({u'}_k^\mu)_0 n_k \delta\!\left(\mathcal{U}_k^{(\mu)}(x,0;t')_0-x'\right)
\label{t=0scf}\end{align}
and
\begin{align}
N\int_{-1/2}^{1/2}dxx\bbra s(0,0)s(x,t)\kket^\ms{c}_N
&=\sum_{j,k} \int_{-1/2}^{1/2}dxx\int d\la \left[n_j^{\dr;\la} \Gamma_j^{(\la)}\right] \!(0)_0  \nn\\
&\qquad\qquad\qquad\qquad\times
    (R^{-\ms{T}}_{jk})^\la_{\;\,\mu}(u_k^\mu)_0 n_k \delta\!\left(\mathcal{U}_k^{(\mu)}(0,0;t)_0-x\right)
     \nn\\
&=\sum_{j,k} \int d\la \left[n_j^{\dr;\la} \Gamma_j^{(\la)}\right] \!(0)_0  \nn\\
&\qquad\qquad\qquad\qquad\times
    (R^{-\ms{T}}_{jk})^\la_{\;\,\mu}(0)_0 n_k\,\mathcal{U}_k^{(\mu)}(0,0;t)_0,  
\end{align}
which leads to
\begin{equation}
N\int_{-1/2}^{1/2}dxx\bbra s(0,0)s(x,0)\kket^\ms{c}_N=0,
\qquad
N\int_{-1/2}^{1/2}dxx\bbra s(0,0)s(x,\tau)\kket^\ms{c}_N=0.
\label{0-1,2}\end{equation}
The second equality in Eq.s\,\eqref{0-1,2} is obtained from the symmetric relations 
$[n_j^{\dr;-\la} \Gamma_j^{(-\la)}](x,t)=[n_j^{\dr;\la} \Gamma_j^{(\la)}](x,t)$,
$(R^\ms{T}_{jk})^{-\la}_{\;\,-\mu}(x,t)=(R^{\ms{T}}_{jk})^\la_{\;\,\mu}(x,t)$
and
\begin{equation}
\mathcal{U}_k^{(-\mu)}(0,0;\tau)=-\mathcal{U}_k^{(\mu)}(0,0;\tau).
\label{anti}\end{equation}
The relation \eqref{anti} is seen from rescaling the space element as $d\hat{x}_j^{(\la)}(t)=\rho_j^{\ms{t};\la}(x,t)dx$ and rewriting Eq.\,\eqref{characteristics} as \cite{Doyon lec.}  
\begin{equation}
\hat{x}_j^{(\la)}(t)-\hat{u}_j^{(\la)}(0)=\hat{v}_j(\la)t,
\label{characteristics 2}\end{equation}
from which we have
$\hat{x}_j^{(-\la)}(t)=\hat{v}_j(-\la)t=-\hat{v}_j(\la)t=-\hat{x}_j^{(\la)}(t)$ at $\hat{u}_j^{(\la)}(0)=\hat{u}_j^{(-\la)}(0)=0$. This is equivalent to Eq.\,\eqref{anti} due to the relation $\rho_j^{\ms{t};-\la}(x,t)=\rho_j^{\ms{t};\la}(x,t)$.  
Let us rewrite Eq.\,\eqref{t=0scf} identically as
\begin{align}
\bbra s(x,0)s(x',t')\kket^\ms{c}_N
&=N^{-1}\sum_{j,k} \int d\la
    \left[n_j^{\dr;\la} \Gamma_j^{(\la)}\right]\!(x)_0
    (R^{-\ms{T}}_{jk})^\la_{\;\,\mu}({u'}_k^\mu)_0 n_k\delta\!\left(x-{u'}_k^\mu\right),
\end{align}
which leads to
\begin{align}
N\int_{-1/2}^{1/2}dxx\bbra s(x,0)s(0,\tau)\kket^\ms{c}_N
&=\sum_{j,k} \int_{-1/2}^{1/2}dxx\int d\la \left[n_j^{\dr;\la} \Gamma_j^{(\la)}\right]
   (x)_0  \nn\\
&\qquad\qquad\qquad\qquad\times
    (R^{-\ms{T}}_{jk})^\la_{\;\,\mu}\!\left(u_k^{(\mu)}(0,\tau)_0\right)_0 n_k \delta\!\left(x-u_k^{(\mu)}(0,\tau)_0\right)   \nn\\
&=\sum_{j,k} \int d\la \left[n_j^{\dr;\la} \Gamma_j^{(\la)}\right]
   \!\left(u_k^{(\mu)}(0,\tau)_0\right)_0  \nn\\
&\qquad\qquad\qquad\qquad\times
    (R^{-\ms{T}}_{jk})^\la_{\;\,\mu}\!\left(u_k^{(\mu)}(0,\tau)_0\right)_0 n_k  u_k^{(\mu)}(0,\tau)_0.   
\end{align}  
We have $\hat{u}_k^{(-\mu)}(0)=-\hat{u}_k^{(\mu)}(0)$ from Eq.\,\eqref{characteristics 2} at $\hat{x}_k^{(\mu)}(t)=\hat{x}_k^{(-\mu)}(t)=0$, or equivalently $u_k^{(-\mu)}(0,t)=-u_k^{(\mu)}(0,t)$.
We also have the symmetric field \eqref{m. field}, $h_0(-x)=h_(x)$, from which it follows that $[n_j^{\dr;-\la} \Gamma_j^{(-\la)}](u_k^{(-\mu)}(0,\tau)_0)_0=[n_j^{\dr;\la} \Gamma_j^{(\la)}](u_k^{(\mu)}(0,\tau)_0)_0$ and
\begin{equation}
N\int_{-1/2}^{1/2}dxx\bbra s(x,0)s(0,\tau)\kket^\ms{c}_N=0.
\end{equation}
From this and Eq.s\,\eqref{0-1,2}, we obtain the relation \eqref{sigma}.

\section{The $\beta\to0$ limit of $\s(\beta)$ }\label{h.t.limit}\label{divergence}
Here we calculate the high temperature limit of the spin conductivity \eqref{sigma}.
At first, we note that the solutions to the TBA equations $\eta_j^{(\la)}(x)$ \eqref{TBA} are independent of the spectral parameter $\la$ in the limit $\beta\to0$. 
With the ratio $\beta h$ kept finite, it is given as \cite{Takahashi 99}
\begin{align}
\eta_j^{(\la)}(x)=
\begin{dcases}
&\!\left\{\frac{\sinh(\n_{j+1}\beta h(x))}{\sinh(y_r\beta h(x))}\right\}^2-1
                 =\eta_j+O(\beta^2),  \\
&\qquad \eta_j:=\frac{n_j \n_{j+2}}{y_r^2} 
  \qquad \for\quad m_r \le j <m_{r+1}, \quad 1\le j \le m_\alpha-2,   \\
&\!\left(\frac{\sinh (n_{m_\alpha-1}\beta h(x))}
                                     {\sinh(n_{m_\alpha}\beta h(x))}\right)^{1-2\delta_{j,m_\alpha}}e^{y_\alpha\beta h(x)}  
                 =\eta_j\{1+y_\alpha\beta h(x)\}  +O(\beta^2),\\
&\qquad \eta_j:=\left(\frac{n_{m_\alpha-1}}{n_{m_\alpha}}\right)^{1-2\delta_{j,m_\alpha}}
  \qquad \for\quad j=m_\alpha-1,\,m_\alpha,
\end{dcases}
\label{etalim}\end{align}
where the modified TS numbers $\{\n_j\}_{j=1}^{m_\alpha}$ and the sequence of numbers $\{y_r\}_{r=-1}^\alpha$ are defined in Appendix \ref{TS no.}.
From Eq.\,\eqref{etalim}, it follows that one-particle magnetization $n_j^\dr(x)$ \eqref{1 mag} appears only on the final boundary $n_{m_\alpha-1}$- and $n_{m_\alpha}$-strings in the limit $\beta\to0$ as
\begin{equation}
n^\dr_j(x)=\begin{dcases}
\frac{y_\alpha}{2}+ O(\beta ) & \for \quad j = m_{\alpha-1}, \; m_\alpha \\
O(\beta ) & \text{otherwise}
\end{dcases}
\label{nlim}\end{equation}
and it is only the driving term that remains in the $\beta\to0$ limit of the quantity $\mathcal{E}_j^{(\la)}(x,t')_0$ \eqref{version 1} as
\begin{equation}
\mathcal{E}_j^{(\la)}(x,t')_0=\mathcal{E}_{0;j}^{(\la)}(x,t')_0+O(\beta).
\label{d.only}\end{equation}
Second, we integrate the driving term $\mathfrak{S}_{0;j}^{(\la)}(x,t')_0=d_j^{(\la)}(x,t')_0+{d'}_j^{(\la)}(x,t')_0$ \eqref{d. term} as
\begin{align}
&\int_{-1/2}^{1/2}dx'x' \int d\la 
    \left[n_j^{\dr;\la} \Gamma_j^{(\la)} \mathfrak{S}_{0;j}^{(\la)}\right](0,\tau)_0 \nn\\
&=\varsigma_j \int d\la 
   (-1)^{\Theta\!\left(u_j^{(\la)}(0,\tau)_0\right)}
   \!\left[\frac{(n_j^{\dr;\la})^2}{\left(1+\eta_j^{(\la)}\right)\left(1+\left(\eta_j^{(\la)}\right)^{-1}\right)} \right]
   \!\left(u_j^{(\la)}(0,\tau)_0\right)_0   \nn\\
&\quad \times \sum_k\int_{-1/2}^{1/2}dx'x'
  \Biggl\{\left[(T^\dr_{jk})^\la_{\;\,\mu}\Gamma_k^{(\mu)}
           \frac{\del \ln\eta_k^{(\mu)}}{\del u_k^{(\mu)}(x',\tau)_0}\right]\!\left(u_k^{(\mu)}(x',\tau)_0\right)_0
  (-1)^{\Theta\!\left(u_k^{(\mu)}(x',\tau)_0\right)}\Theta(-x')  \nn\\
&\qquad -
  \left[(T^\dr_{jk})^\la_{\;\,\mu}\Gamma_k^{(\mu)}
           \frac{\del \ln\eta_k^{(\mu)}}{\del x'}\right]\!(x')_0
  (-1)^{\Theta(x')}\Theta\!\left(u_j^{(\la)}(0,\tau)_0-x'\right) \Biggr\}.
\label{d. intg}\end{align}  
This is seen from transforming the factors $\Gamma_j^{(\la)}(x)_0$ \eqref{density}, $\mathcal{W}_j^{(\la)}(x)_0$ \eqref{version 2} and $n_j^{\dr;\la}(x)_0$ \eqref{1 mag} as
\begin{align}
&\left[\Gamma_j^{(\la)}\mathcal{W}_j^{(\la)}\right]\!(x)_0
=
(-1)^{\Theta(x)}4\beta h_0 \left[\frac{n_j^{\dr;\la}}{\left(1+\eta_j^{(\la)}\right)\left(1+\left(\eta_j^{(\la)}\right)^{-1}\right)}\right]\!(x)_0,  \nn\\
&n_j^{\dr;\la}(x)_0
=
\frac{\del_x \ln\eta_j^{(\la)}(x)_0}{2\beta\del_x h_0(x)}
=
(-1)^{\Theta(x)}\frac{\del_x \ln\eta_j^{(\la)}(x)_0}{4\beta h_0}.
\end{align}
with the gradient of the magnetic field \eqref{m. field}, $\del_xh_0(x)=(-1)^{\Theta(x)}2h_0$.
Using Eq.s\,\eqref{etalim} and \eqref{nlim}, we have the sum over the string number of the $\beta\to0$ limit of Eq.\,\eqref{d. intg} as
\begin{align}
&\lim_{\beta\to0}\sum_j\int_{-1/2}^{1/2}dx'x' \int d\la 
    \left[n_j^{\dr;\la} \Gamma_j^{(\la)} \mathfrak{S}_{0;j}^{(\la)}\right](0,\tau)_0 \nn\\
&= -\frac{\varsigma_{m_\alpha-1}y_\alpha^2}{2(1+\eta_{m_\alpha-1})(1+\eta_{m_\alpha-1}^{-1})}
    \sum_k \ln\eta_k \int_{-\infty}^\infty d\la \int_{-\infty}^\infty d\mu 
   (-1)^{\Theta\!\left(u_{m_\alpha-1}^{(\la)}(0,\tau)_0\right)} T^\dr_{m_\alpha-1,k}(\la-\mu)\Gamma_k(\mu)  \nn\\
&\quad \times \int_{-1/2}^{1/2}dx'  (-1)^{\Theta\!\left(u_k^{(\mu)}(x',\tau)_0\right)}\Theta(-x')  \nn\\  
&= (-1)^{\Theta(A)}\varsigma_{m_\alpha-1}n_{m_\alpha-1}n_{m_\alpha}
    \sum_k \ln\eta_k \int^{\infty}_0 d\la \int_{-\infty}^\infty d\mu 
    T^\dr_{m_\alpha-1,k}(\la-\mu)\Gamma_k(\mu) v^\mr{eff}_k(\mu)\tau,
\label{cal.}\end{align}
where all physical excitations of strings are independent of $x$ in the limit $\beta\to0$, which let the space coordinate disappear.
To obtain the second expression of Eq.\,\eqref{cal.}, we note at first that one-particle magnetization Eq.\,\eqref{nlim} is used and the integration by parts is performed;
second, the derivative with respect to the initial position $u_j^{(\la)}(x,t)_0$ is replaced by the derivative with respect to $x$ since differentiating the both sides of Eq.\,\eqref{characteristics} with respect to $x$ yields $\del_xu_j^\la=1$ at a fixed $t$;
third, the total derivative terms do not remain if the following conditions are satisfied: 
\begin{equation}
u_k^{(\mu)}(-1/2,\tau)<0 \quad \mr{and} \quad |u_{m_\alpha-1}^{(\la)}(0,\tau)|<1/2.
\label{flight}\end{equation}
These conditions are satisfied by assuming the displacement of the physical excitation to be shorter than a half of the system size $N$ in time $\mathcal{T}=N \tau$;
fourth, the parts from the second driving term ${d'}_j^{(\la)}(x,t)$ are also canceled due to symmetric relations in the spaces of $x$ and $\la$ such as $u_{m_\alpha-1}^{(-\la)}(0,\tau)=-u_{m_\alpha-1}^{(\la)}(0,\tau)$, which is obtained as same as Eq.\,\eqref{anti}, $T^\dr_{jk}(-\la)=T^\dr_{jk}(\la)$ and so on.
The third expression of Eq.\,\eqref{cal.} is obtained from the symmetric relations and the $\beta\to0$ limit of Eq.\,\eqref{characteristics 2}, 
$
\lim_{\beta\to0}u_k^{(\mu)}(x',\tau)=x'-v_k^\mr{eff}(\mu)\tau 
$
with the parity of the effective velocity determined by Eq.\eqref{aj} and \eqref{ev}; for example,
$v_j^\mr{eff}(\la) \le 0$ for $\la \ge 0$ and $A<0$.
Using Eq.\,\eqref{cal.} with Eq.\,\eqref{lrscf} for $S(x,x';t)$, \eqref{eos} for $v_j^\mr{eff}(\la)$ and \eqref{density} for $\Gamma_j(\la)$, we find that the spin conductivity $\s(\beta)$ \eqref{sigma} is proportional to $\beta$ in the high temperature limit as
\begin{align}
\s_0:=\lim_{\beta\to0}\frac{\s(\beta)}{\beta}
&=\varsigma_{m_\alpha-1}|A|\frac{n_{m_\alpha-1}n_{m_\alpha}}{4\pi}\sum_j 
    \frac{\varsigma_j\ln\eta_j}{(1+\eta_j)(1+\eta_j^{-1})}  \nn\\
&\quad\times     \int_{-\infty}^\infty \! d\la \,
    \left\{\del_{\beta A}\ln\eta_j^{(m_\alpha-1)}(\la)-\del_{\beta A}\ln\eta_j^{(m_{\alpha-1})}(\la)\right\}
    \del_{\beta A}\ln\eta_j(\la).
\label{sigma b0}\end{align}
Here $\eta_j^{(j_\s)}(\la)$ denotes the solutions to the TBA equations for the integrable XXZ chain with spin-$\s/2$, which satisfies the condition that the number $\s$ is commensurable with the modified TS numbers as $\s+1\in\{\n_j\}_{j=1}^{m_\alpha}$ \cite{Kirillov,Kirillov2,AS,Ae 2024}. The number $j_\s$ is specified by the relation $\s+1=\n_{j_\s}$, and $\ln\eta_j(\la)=\ln\eta_j^{(2)}(\la)$ in Eq.\,\eqref{sigma b0}.
As shown in \cite{Ae 2024}, the $\beta\to0$ limit of the final boundary dressed scattering kernel is given by
\begin{equation}
T^\dr_{m_\alpha-1,j}(\la) = \del_{\beta A}\ln\eta_j^{(m_\alpha-1)}(\la)-\del_{\beta A}\ln\eta_j^{(m_{\alpha-1})}+O(\beta A)
\label{Tdrex}\end{equation} 
and calculated from the following expansion of $\ln\eta_j^{(j_\s)}(\la)$: 
\begin{equation}
\eta^{(j_\s)}_j(\la)
=\begin{dcases}
\left(\frac{\n_{j+1}}{y_r}\right)^2 
\left(1-\frac{\n_{j+2}\beta A}{y_r\n_{j+1}}\sum_{s=1}^\s\left(1-\frac{s}{\s+1}\right)s\varDelta a^{(j_\s)}_{j,s}(\la)\right)-1+O\!\left((\beta A)^2\right)\\
          \qquad \for \;\; m_r \le j < m_{r+1}, \quad 1\le j \le m_\alpha-2,  \\ 
\frac{y_\alpha}{y_\alpha-n_j}\left(1-\frac{\beta A}{y_\alpha-n_j}\sum_{s=1}^\s\left(1-\frac{s}{\s+1}\right)s\varDelta a^{(j_\s)}_{j,s}(\la)\right)-1 
+O\!\left((\beta A)^2\right) \\
          \qquad \for \;\; j = m_\alpha-1,\,m_\alpha,
\end{dcases}
\end{equation}
with
\begin{align}
\varDelta a^{(j_\s)}_{j,s}(\la)&:=a(\la;\tilde{q}_{j_\s}+q_j+2s)
-\Theta(m_\alpha-2-j)\frac{n_j}{\n_{j+2}}a(\la;\tilde{q}_{j_\s}+\tilde{q}_{j+2}+2s),  \nn\\  
a(\la;q)&:=\frac{\theta}{2\pi}\frac{\sin\theta q}{\cosh\theta\la-\cos\theta q},
\label{a's def}\end{align}
where the sequence of numbers $\{\tilde{q}_j\}$ is defined in Appendix \ref{TS no.}.
To make sure that this paper is self-contained, we list the result obtained in \cite{Ae 2024} as\footnote
{
We made a typo in Eq.\,(3.14) of Ref.\,\cite{Ae 2024} that $\varDelta a^{(m_{\alpha-1})}_{j,s} (\la)$ was wrongly written as $\varDelta a^{(m_{\alpha})}_{j,s} (\la)$. This has been corrected in Eq.\,\eqref{Tdr}. The definition Eq.\,\eqref{a's def} is changed from the original one in Eq\,(3.12) of \cite{Ae 2024}.  
} 
\begin{align}
T^\dr_{m_\alpha-1,j}(\la) 
&\!=\!
-\frac{\n_{j+1}}{y_r n_j }
\!\left\{
\sum_{s=1}^{n_{m_\alpha-1}-1}\!\left(1-\!\frac{s}{n_{m_\alpha-1}}\right) \!s \varDelta a^{(m_\alpha-1)}_{j,s} (\la)
-\!
\sum_{s=1}^{n_{m_\alpha}-1}\!\left(1-\!\frac{s}{n_{m_\alpha}}\right)\!s \varDelta a^{(m_{\alpha-1})}_{j,s} (\la)
\!\right\}  
\nn\\
&=
\frac{\n_{j+1}y_\alpha}{\n_{j+2}n_{m_\alpha-1}n_{m_\alpha}}K_{m_\alpha-1,j}^\dr(\la)
\quad  \for \;\; m_r \le j < m_{r+1},  \;\;   1 \le j \le m_\alpha-2,  \nn\\
T^\dr_{m_\alpha-1,j}(\la) 
&=
-\frac{y_\alpha}{n_{m_\alpha-1}n_\alpha}
\!\left\{
\sum_{s=1}^{n_{m_\alpha-1}-1}\!\left(1-\!\frac{s}{n_{m_\alpha-1}}\right) \!s \varDelta a^{(m_\alpha-1)}_{j,s} (\la)
-\!
\sum_{s=1}^{n_{m_\alpha}-1}\!\left(1-\!\frac{s}{n_{m_\alpha}}\right)\!s \varDelta a^{(m_{\alpha-1})}_{j,s} (\la)
\!\right\}  \nn\\
&=
(-1)^{\delta_{j,m_\alpha}}\frac{y_\alpha}{n_{m_\alpha-1}n_{m_\alpha}}K_{m_\alpha-1,j}^\dr(\la)
\quad  \for \quad j=m_{\alpha}-1, \; m_\alpha \; (r=\alpha-1)
\label{Tdr}\end{align}
and
\begin{align}
K^\dr_{m_\alpha-1,j}(\la) 
&=
\sum_{s=1}^{n_j-1}
\left(\frac{2s(n_j-s)}{n_j}+y_r\right)
a(\la; q_{m_\alpha}+q_j+2s)
\nn\\
&\quad
+\sum_{s=1}^{y_r}\frac{s^2}{y_r}
\bigg\{
a(\la; q_{m_\alpha}+\tilde{q}_{j+2}+2s)
+a(\la; q_{m_\alpha}-\tilde{q}_{j+2}-2s)
\bigg\} 
\nn\\
\for & 
\quad m_r\le j<m_{r+1}  \;(r \le \alpha-2)  
\;\; \textup{and} \;\; 
j=m_{\alpha-1},  
\nn\\
K^\dr_{m_\alpha-1,j}(\la) 
&=
\frac{2n_{m_\alpha}}{n_j}\sum_{s=1}^{\left\lfloor\frac{n_{j-1}-1}{2}\right\rfloor}
(n_{j-1}-2s)a(\la; q_{j-1}+2s)
\nn\\
&\quad
+\!\sum_{s=1}^{n_{m_\alpha}}
\!\Bigg\{\!\left(2s+n_{m_\alpha} \!-\frac{\n_{j+2}s^2}{n_jn_{m_\alpha}} \right)
\! a(\la; q_{j+1}+2s)
+
\! \frac{s^2}{n_{m_\alpha}}  
a(\la; q_{m_\alpha} \!+\tilde{q}_{j+2}+2s)
\!\Bigg\}
\nn\\
\for & 
\quad m_{\alpha-1} < j \le m_\alpha-2,  
\nn\\
K^\dr_{m_\alpha-1,m_\alpha-1}(\la)
&=-K^\dr_{m_\alpha-1,m_\alpha}(\la)  \nn\\
&=n_{m_\alpha}\sum_{s=0}^{\left\lfloor\frac{y_\alpha}{2}\right\rfloor-n_{m_\alpha}}
\left(\frac{n_{m_\alpha}+2s}{n_{m_\alpha-1}}-1\right)a(\la;2q_{m_\alpha}-2s)  \nn\\
&\quad
+\sum_{s=1}^{n_{m_\alpha}-1}s\left(2-\frac{y_\alpha s}{n_{m_\alpha-1}n_{m_\alpha}}\right)a(\la;2s),
\end{align}
where $\lfloor z \rfloor$ denotes the largest integer less than or equal to $z$.
The high temperature proportionality constant \eqref{sigma b0}, $\s_0=\sum_j\s_j$, is written as
\begin{align}
\s_j
&:=|A|\frac{n_{m_\alpha-1}n_{m_\alpha}}{4\pi}\sum_j 
    \frac{\ln\eta_j}{(1+\eta_j)(1+\eta_j^{-1})} \int_{-\infty}^\infty 
    \rho^\ms{t}_{m_\alpha-1,j}(\la)\rho^\ms{t}_j(\la) d\la \nn\\
&=\begin{dcases}
 -\varsigma_{m_\alpha-1} \varsigma_j\frac{|A|}{8\pi}\frac{y_\alpha y_r}{\n_{j+1}^2}\ln\frac{n_j\n_{j+2}}{y_r^2} 
    \int^\infty_{-\infty} K^\dr_{m_\alpha-1,j}(\la)\varDelta a_j(\la) d\la      \\
          \qquad \for \;\; m_r \le j < m_{r+1}, \quad 1\le j \le m_\alpha-2,  \\
 -\varsigma_{m_\alpha-1} \varsigma_j\frac{|A|}{8\pi}
 \ln\left(\frac{n_{m_\alpha-1}}{n_{m_\alpha}}\right)^{1-2\delta_{j,m_\alpha}} 
 \int_{-\infty}^\infty K^\dr_{m_\alpha-1,j}(\la)\varDelta a_j(\la) d\la  \\
          \qquad \for \;\; j = m_\alpha-1,\,m_\alpha.
\end{dcases}
\label{for Fig}\end{align}
In the second expression of Eq.\,\eqref{for Fig}, $\rho^\ms{t}_j(\la)=\varsigma_j\del_{\beta A}\ln\eta_j(\la)$ is the density of vacancies for the one-particle excitation from the state where all spins are up. The same for the two-particle excitation \cite{form factor,ND diffusion} is defined by $\rho_{m_\alpha-1,k}^\ms{t}(\la):=\varsigma_{m_\alpha-1}T^\dr_{m_\alpha-1,j}(\la)$ from the identity for $T^\dr_{m_\alpha-1,j}(\la)$ Eq.\,\eqref{Tdrex}. As explained in \cite{Ae 2024}, the dressed scattering kernel $T^\dr_{m_\alpha-1,j}(\la)$ represents the rescaled energy of the two-particle excitation with the scaling factor $A=-2\pi J\sin\theta/\theta$ \eqref{aj}.
In the last expression of Eq.\,\eqref{for Fig}, $\varDelta a_j(\la):=\varDelta a_{j,1}^{(2)}(\la)$.

We have evaluated the constant \eqref{sigma b0} for the XXX chain $\s^\mr{xxx}_0=\sum_j\s^\mr{xxx}_j$. As we prove below, $\s^\mr{xxx}_0$ diverges in the scale of $(\ln p_0)^2$ for which the number $p_0$ is infinite at the isotropic point $\Delta=\cos\pi/p0=1$ \eqref{anisotropy} and represents the one-particle magnetization as $p_0=n_{p_0-1}+n_{p_0}$ \eqref{nlim} at $\Delta=1$ in the limit $\beta\to0$: 
\begin{proof}
At first we note that the string length $n$ for the XXX chain is arbitrary in the thermodynamic limit as
$1\le n \le p_0=N/2=\infty$ \cite{Takahashi XXX}.
The Fourier transform of Eq.\,\eqref{a's def} is given by
\begin{equation}
\widehat{a}(\omega;n)
=\frac{\sinh(p_0-n)\omega}{\sinh p_0\omega }
=e^{-n|\omega|}
\end{equation}
at $p_0=\infty$. Using this, $\s^\mr{xxx}_0$ is rewritten as
\begin{align}
\s^\mr{xxx}_0
&=\frac{p_0|J|}{4}\sum_{n=1}^{p_0-2}\frac{\ln[n(n+2)]}{(n+1)^2}
  \int_{-\infty}^\infty\frac{d\omega}{2\pi}
  \widehat{K^\dr_{p_0-1},n}(\omega) \widehat{\varDelta a_n}(\omega) \nn\\
&=\frac{p_0|J|}{4}\sum_{n=1}^{p_0-2}\frac{\ln[n(n+2)]}{(n+1)^2}  
    \int_{-\infty}^\infty\frac{d\omega}{2\pi}  \nn\\
&\quad\times    \Biggl[\frac{2}{n}\sum_{s=1}^{\lfloor\frac{n-2}{2}\rfloor}(n-1-2s)
           \frac{\sinh(n-1-2s)\omega}{\sinh p_0\omega }  +\left(2-\frac{2}{n}\right) \frac{\sinh(n-1)\omega}{\sinh p_0\omega }
           +\frac{\sinh(n+1)\omega}{\sinh p_0\omega } \Biggr]   \nn\\
&\quad \times\left[
\frac{\sinh(p_0-n)\omega}{\sinh p_0\omega }-\frac{n}{n+2}\frac{\sinh(p_0-n-2)\omega}{\sinh p_0\omega }
\right]  \nn\\
&=\frac{p_0|J|}{4}\sum_{n=1}^{p_0-2}\frac{\ln[n(n+2)]}{(n+1)^2}  \nn\\
&\quad \times  \int_{-\infty}^\infty\frac{d\omega}{2\pi} 
   \Biggl[\frac{2}{n}\sum_{s=1}^{\lfloor\frac{n-2}{2}\rfloor}(n-1-2s)\frac{e^{(n-1-2s)|\omega|}-e^{-(n-1-2s)|\omega|}}{e^{p_0|\omega|}}  \nn\\ 
&\quad  +\left(2-\frac{2}{n}\right) \frac{e^{(n-1)|\omega|}-e^{-(n-1)|\omega|} }{e^{p_0|\omega|}}
           +\frac{e^{(n+1)|\omega|}-e^{-(n+1)|\omega|} }{e^{p_0|\omega|}} \Biggr]   \nn\\
&\quad \times\left[
e^{-n|\omega|}- \frac{n}{n+2} e^{-(n+2)|\omega|} 
\right]  +O(\ln p_0)\nn\\
&=\frac{|J|}{\pi}\sum_{n=1}^{p_0-2}\frac{\ln[n(n+2)]}{(n+1)^2(n+2)} 
\left[\frac{2}{n}\sum_{s=1}^{\lfloor\frac{n-2}{2}\rfloor}(n-1-2s)^2+3n-3+\frac{2}{n}\right]+O(\ln p_0).
\label{XXX scale}\end{align}
This value is scaled to $O((\ln p_0)^2)$ with respect to $p_0$.
\end{proof}

\begin{figure}[H]
    \centering
    \includegraphics[width=1\columnwidth]{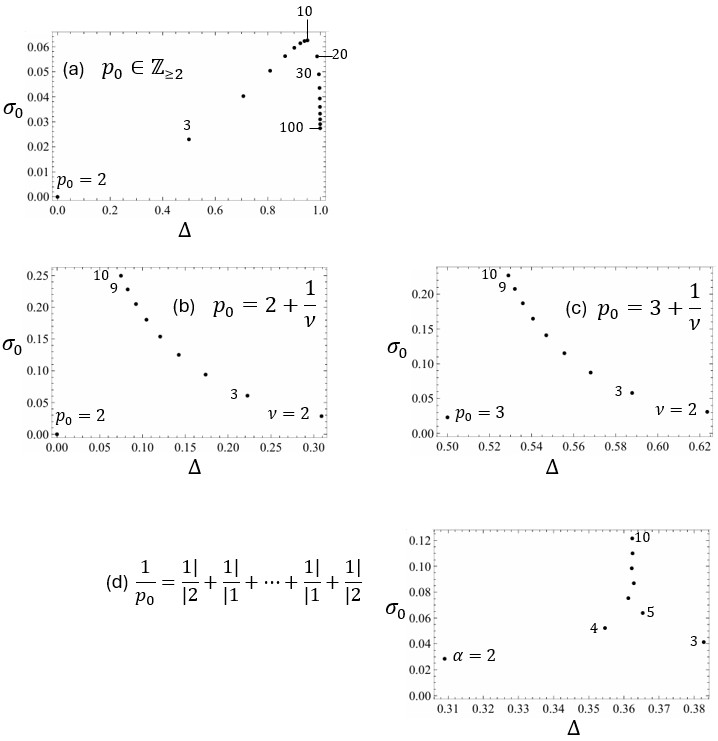}
\caption{$\s_0$ calculated for various numbers of the anisotropy parameter; (a) $p_0=2, 3, \cdots, 10, 20, \cdots, 100$, (b) $p_0=2+1/\nu$ with $\nu=2, 3, \cdots,10$, (c) $p_0=3+1/\nu$ with $\nu=2, 3, \cdots,10$ and
(d) $p_0$ is given by the continued fraction\\
\centerline{$\dfrac{1}{p_0}=\dfrac{1|}{|2}+\dfrac{1|}{|1}+\cdots+\dfrac{1|}{|1}+\dfrac{1|}{|2}
=\dfrac{1}{2+\dfrac{1}{1+\dfrac{1}{\dfrac{\ddots}{1+\dfrac{1}{2}}}}}$}
with length $\alpha=2, 3, \cdots, 10$. $1/p_0=1|/|2+1|/|2$ for $\alpha=2$, $p_0=1|/|2+1|/|1+1|/|2$ for $\alpha=3$ and so on. At $\alpha=\infty$, the golden number $(1+\sqrt{5})/2$ appears as $p_0=1+(1+\sqrt{5})/2$.
The points ($\bullet$) indicate the values for $\s_0=\sum_j\s_j$ \eqref{for Fig} with the coupling constant $J=1$ for the factor $A=-2\pi J\sin\theta/\theta$ \eqref{aj}.}  
\label{Fig.s}  
\end{figure}

In Fig.s\,\ref{Fig.s}, we show $\s_0$ for the XXZ chain as a function of the anisotropy parameter $\Delta$, which approaches the (a) isotropic point $\Delta=1$, (b) free fermion point $\Delta=0$, (c) $\Delta=0.5$ and the point indicated by (d) $p_0=1+(1+\sqrt{5})/2$ where $(1+\sqrt{5})/2$ is the golden number.
As shown in Fig.\,\ref{Fig.s}-(a), $\s_0$ decays to zero when $p_0$ approaches the isotropic point, while $\s_0^\mr{xxx}$ diverges. On the contrary, $\s_0$ increases infinitely when $p_0$ approaches any rational number other than $p_0=\infty$, while $\s_0$ is a finite value when $p_0$ is exactly on the rational numbers. This behavior is seen in Fig.\,\ref{Fig.s}-(b) and (c), in the former of which the value at the free fermion point $(p_0=2)$ is zero.
$\s_0$ also increases infinitely when $p_0$ approaches irrational numbers as Fig.\,\ref{Fig.s}-(c), while it remains a problem whether the value is infinite or not when $p_0$ is exactly on irrational numbers.
The spin transport is superdiffusive in the case where $\s_0$ is infinitely large. The divergence of $\s_0^\mr{xxx}$ is indeed an evidence of the superdiffusion at the isotropic point \cite{ND diffusion,ND diffusion (0),Gopalakrishnan2}.

\section{Summary and discussion}
In the critical regime, we have set up the spin-1/2 XXZ chain with a gradient of magnetization concentration.
By applying an appropriate magnetic field, the slope of the concentration goes down constantly from the center to both ends of a spin chain which runs an infinite length $N$.    .
The magnetization starts to change after removing the filed and the $1/N$-scaled long-range spin correlation is generated in the non-equilibrium state. We have found that the fluctuation of the spin current, or the spin conductivity $\s(\beta)$ is driven by the long-range spin correlation. In the high temperature limit $\beta\to0$, $\s(\beta)$ is proportional to $\beta$ and the constant $\s(\beta)/\beta$ diverges in the case where one-particle magnetization is infinitely large in the critical regime plus at the isotropic point. This divergence informs that spin transport can be superdiffusive when it is driven by the long-range spin correlation.

Let us discuss the dynamic scaling in spin transport at the isotropic point. Using the GHD theory, its dynamical relationship at infinite temperature is consistently shown to be the same as $x\sim t^{2/3}$ \cite{Gopalakrishnan2}, which is the universal character obtained by Kardar, Parisi and Zhang (KPZ) for surface growth phenomena \cite{KPZ}. Numerical studies also suggest that the XXX chain belongs to the KPZ universality class \cite{KPZ-XXX}.
On the other hand, a recent investigation \cite{KPZ no XXX} indicates that the transport at the isotropic point cannot be explained by the KPZ type, which has also been supported by quantum simulations \cite{KPZ no XXX 2}.  
Through the present study employing the BMFT, we are now provided with an analytic approach in this regard; as suggested by the direct analogy with electric conductivity which is expressed by the concentrations of electrons and holes and by their mobilities, $\s_0(\beta)$ may be expressed as 
\begin{equation}
\s(\beta)=\sqrt{\overline{\varDelta s^2}}\frac{v}{\del_\mr{x}h(\mr{x}/N)},
\end{equation}
where $\overline{\varDelta s^2}$ is the average fluctuation of spin density and $v$ is its displacement velocity. When the fluctuation of the spin current is driven by the long-range correlation \eqref{lrscf}, $\overline{\varDelta s^2}$ is given by
\begin{equation}
\overline{\varDelta s^2}=\int_{-1/2}^{1/2} dx S(0,x';\tau).
\end{equation}
At the isotropic point $p_0=N/2=\infty$, this is scaled as $\overline{\varDelta s^2}\sim O((\ln N)^2/N)$ in the limit $\beta\to0$, which leads to $v/\beta\sim O(\ln N/N^{1/2})$ in conjunction with Eq.\,\eqref{XXX scale}, $\s_0\sim O((\ln N)^2)$.
Let $\mathcal{T}$ be the time in which the spin is transferred from the center to both ends of its chain
Then we have a dynamical relationship $\mathcal{T}\sim N^{3/2}/\ln N$.
In this way the spin transport is enhanced beyond the KPZ dynamical relationship by a logarithmic correction in the XXX chain.


\begin{appendix}
\section{TS numbers}\label{TS no.}
Starting from the anisotropy parameter $\theta$, the series of numbers
$\{p_r\}_{r=0}^{\alpha+1}, \; \{\nu\}_{r=1}^{\alpha+1}, \; \{m_r\}_{r=0}^{\alpha+1}$ and $\{y_r\}_{r=-1}^\alpha$ are defined as \cite{TS 72,Takahashi 99} 
\begin{align}
&p_0=\frac{\pi}{\theta}, \quad p_1=1, \quad \nu_r=\left\lfloor\frac{p_{r-1}}{p_r}\right\rfloor, 
\quad p_r=p_{r-2}-\nu_{r-1}p_{r-1}, \nn\\
&p_{\alpha+1}=0, \quad \nu_{\alpha+1}=\infty, \nn\\
&m_0=0, \quad m_r=\sum_{k=1}^r \nu_k, \quad m_{\alpha+1}=\infty, \nn\\
&y_{-1}=0, \quad y_0=1, \quad y_1=\nu_1 \quad \mathrm{and} \quad y_r=y_{r-2}+\nu_r y_{r-1}.  
\label{p, m, y}\end{align}
The TS numbers $\{n_j\}_{j=1}^{m_\alpha}$, the associated parities $\{v_j\}_{j=1}^{m_\alpha}$ 
and numbers $\{q_j\}_{j=1}^{m_\alpha}$ are determined as follows:
\begin{align}
&  n_j=y_{r-1}+(j-m_r)y_r  \qquad (m_r \le  j < m_{r+1}), \nn\\
&  n_{m_\alpha}=y_{\alpha-1},  \nn\\
&  v_{m_1}=-1, \quad v_j=(-1)^{\lfloor(n_j-1)/p_0\rfloor} \qquad (j \ne m_1)  \nn\\
&  \mathrm{and} \;\; q_j = (-1)^r(p_r-(j-m_r)p_{r+1})  \nn\\
&     \qquad\quad \equiv \frac{1+v_j}{2}p_0-n_j \mod 2p_0   \qquad (m_r \le j < m_{r+1} ). 
\end{align}
The modified TS numbers $\{\tilde{n_j}\}_{j=1}^{m_\alpha}$, the associated parities $\{\tilde{v}_j\}_{j=1}^{m_\alpha}$ and numbers $\{\tilde{q_j}\}_{j=1}^{m_\alpha}$ are defined as \cite{KSS,AS,Ae 2024}
\begin{align}
&\tilde{n}_j=y_{r-1}+(j-m_r)y_r \qquad (m_r < j \le m_{r+1}), \nn\\ 
&\tilde{v}_j=(-1)^{\lfloor(\n_j-1)/p_0\rfloor}  \nn\\
&\mathrm{and} \;\; \tilde{q}_j= (-1)^r(p_r-(j-m_r)p_{r+1})  \nn\\
&\qquad\quad \equiv \frac{1+\tilde{v}_j}{2}p_0-\n_j \mod 2p_0   \qquad (m_r < j \le m_{r+1} ).
\label{modified TS no.}\end{align}
Here the identities $\n_j=n_j$, $\tilde{v}_j=v_j$ and $\tilde{q}_j=q_j$ hold except for $j=m_r \;(1\le r \le \alpha)$.

\section{Derivation of Eq.\,\eqref{saddle eq.3}}\label{F. Jacobian} 
First, we diagonalize the flux Jacobian $\mathcal{J}(x,t)$ as Eq.\,\eqref{d.Jaco}. Let us differentiate the density of quasiparticle current $[v_j^{\mr{eff};\la}\rho_j^{(\la)}](x)$ with respect to a parameter or variable $p$ which we have introduced in the dressed function formula \eqref{dff}. 
Denoting that $\delta[v_j^{\mr{eff};\la}\rho_j^{(\la)}](x,t)=\del_p[v_j^{\mr{eff};\la}\rho_j^{(\la)}](x,t)\delta p$, we have 
\begin{align}
\delta \left[v_j^{\mr{eff};\la}\rho_j^{(\la)}\right]\!(x,t)
&=-\delta\left[\frac{\vartheta_j^{(\la)}(\del\epsilon_j)^{\dr;\la}}{2\pi\varsigma_j}\right]\!(x,t)  \nn\\
&=-\frac{1}{2\pi}\sum_k \left[(R^{-1}_{jk})^\la_{\;\,\mu}\varsigma_k\left\{\delta\vartheta_k^{(\mu)}\right\}(\del\epsilon_k)^{\dr;\mu}\right](x,t)   \nn\\
&=-\sum_k \left[(R^{-1}_{jk})^\la_{\;\,\mu}v_k^{\mr{eff};\mu}\Gamma_k^{(\mu)}\delta\ln\eta_k^{(\mu)}\right]\!(x,t).     
\label{C1}\end{align}
These transformations are the same as performed in Eq.\,\eqref{transform} and \eqref{cscf 2}.
In another way,
\begin{align}
\delta \left[v_j^{\mr{eff};\la}\rho_j^{(\la)}\right]\!(x,t)
&=\sum_k\int d\mu \frac{\delta\left[v_j^{\mr{eff};\la}\rho_j^{(\la)}\right]\!(x,t)}{\delta\rho_k^{(\mu)}(x,t)}
                          \delta\rho_k^{(\mu)}(x,t)  \nn\\
&=\sum_k \left[(\mathcal{J}_{jk})^\la_{\;\,\mu}\, \delta \left(\vartheta_k^{(\mu)} \rho_k^{\ms{t};\mu}\right)\right]\!(x,t)  \nn\\
&=-\sum_{k,l} \left[(\mathcal{J}_{jk})^\la_{\;\,\mu} (R^{-1}_{kl})^\mu_{\;\,\nu}\,
\Gamma_l^{(\nu)}\delta\ln\eta_l^{(\nu)}\right]\!(x,t),
\label{C3}\end{align}
in the third expression of which we use the definition of $\mathcal{J}(x,t)$ \eqref{Jacobian}.
Comparing Eq.s\! \eqref{C1} and \eqref{C3}, we obtain
\begin{equation}
(\mathcal{J}_{jk})^\la_{\;\,\mu}(x,t)
=\sum_{l}\left[(R^{-1}_{jl})^\la_{\;\,\nu}v_l^{\mr{eff};\nu}(R_{lk})^\nu_{\;\,\mu}\right]\!(x,t),
\end{equation}
which leads to the diagonalization \eqref{d.Jaco}.

Second, the saddle point equations \eqref{saddle} yield
\begin{equation}
n_k\del_tg_k^{(\mu)}(x,t)  d\la
+ \sum_j n_j \int d\la \left[\del_x g_j^{(\la)} (\mathcal{J}_{jk})^\la_{\;\,\mu}\right](x,t) 
- \frac{c'n_k}{N}\delta(x-x')\delta(t-t') d\la
=0,
\label{C5}\end{equation} 
where the delta function is defined by $\delta(x):=\delta_{0\mr{x}}/dx=\delta_{0\mr{x}}N/L$ and $\delta(t)=\delta_{0\mr{t}}/dt$ with the Kronecker delta $\delta_{\mr{a}\mr{b}}$.
Inserting the identity operator $[R^{-1}R](x,t) =1$ into the sum and integration on the l.h.s.\,of Eq.\,\eqref{C5}, we have
\begin{align}
&n_k\del_tg_k^{(\mu)}(x,t)  d\la-\frac{c'n_k}{N}\delta(x-x')\delta(t-t') d\la \nn\\
&+ \sum_{j,l,m} n_l \int d\nu \left[\del_x g_l^{(\nu)}(R^{-1}_{lm})^\nu_{\;\,\gamma}(R_{mj})^\gamma_{\;\,\la}
   (\mathcal{J}_{jk})^\la_{\;\,\mu}\right](x,t) 
=0.
\end{align} 
Acting by operators $R^{\ms{T}}(x,t)$ and $R^{-1}(x,t)$ on the l.h.s.\,and r.h.s.\,of this equation respectively and using the diagonalization \eqref{d.Jaco}, we obtain
\begin{equation}
n_j\left(\del_tg_j^{(\la)}(x,t) +  \left[\del_x g_j^{(\la)}v^\mr{eff:\la}_j\right](x,t) 
-\frac{c'}{N}\delta(x-x')\delta(t-t')\right)   =0,
\end{equation} 
which means Eq.\,\eqref{saddle eq.3}.

\section{Derivation of Eq. \eqref{version} and \eqref{version 1}}\label{App.D}
We follow here the analysis of Subsec.\,5.5 in Ref.\,\cite{BMFT 1}.
We first provide Eq.s \eqref{A.2}-\eqref{reform} to show that Eq.\,\eqref{g's sol} is transformed into the same form as Eq.\,(165) in \cite{BMFT 1}. Readers may skip to Eq.\,\eqref{g's dress}, for this transformation is not used in the present system.

Dressing the derivatives with respect to $x$ of the both sides of Eq.\,\eqref{g's sol}, we have
\begin{align}
\sum_k\left[(R^{-\ms{T}}_{jk})^\la_{\;\;\mu}n_k  \del_x g_k^{(\mu)}\right]\! (x,t_\epsilon)
&=
-N^{-1}c' n^{\dr;\la}_j(x,t_\epsilon) \del_x\delta(x-x')  \nn\\
&=
-N^{-1}c'
\left[\del_x\!\left\{n^{\dr;\la}_j(x,t_\epsilon) \delta(x-x')\right\}-\del_xn^{\dr;\la}_j(x,t_\epsilon) \delta(x-x') \right].
\label{A.2}\end{align}
Using that
$
\sum_{k}\del_x \!\left[(R_{jk}^\ms{T})^\la_{\;\;\mu}n_k^{\dr;\mu}\right]\!(x,t)=\del_x n_j=0
$,
or equivalently
\begin{equation}
\del_x n_j^{\dr;\la}(x,t)
=
-\sum_{k,l}\left[(R_{jk}^\ms{-T})^\la_{\;\;\mu}\del_x(R_{kl}^\ms{T})^\mu_{\;\;\nu} n_l^{\dr;\nu} \right](x,t)
\end{equation}
and multiplying the both sides of Eq.\,\eqref{A.2} by $\sum_j (R^\ms{T}_{mj})^\gamma_{\;\,\la}(x,t_\epsilon)$, we have
\begin{align}
&n_m \del_xg^{(\gamma)}_m (x,t_\epsilon)  \nn\\
&=
-\frac{c'}{N}\sum_{j}
\Bigl[
(R^\ms{T}_{mj})^\gamma_{\;\,\la}(x,t_\epsilon)\del_x\!\left\{n^{\dr;\la}_j (x,t_\epsilon) \delta(x-x')\right\} 
+\del_x(R_{mj}^\ms{T})^\gamma_{\;\;\la}(x,t_\epsilon) n^{\dr;\la}_j (x,t_\epsilon)\delta(x-x')
\Bigr] \nn\\
&=
-N^{-1}c'\sum_{j}
\del_{x'}\!\left\{(R_{mj}^\ms{T})^\gamma_{\;\;\la}(x',t_\epsilon)n_j^{\dr;\la}(x',t_\epsilon)\delta(x-x')\right\}.
\end{align}
Integrating the both sides of this equality over $[-1/2,x]$ and using the first condition in Eq.s\,\eqref{saddle eq.2}, we have
\begin{equation}
n_j g^{(\la)}_j (x,t_\epsilon)
=
-N^{-1}c'\sum_{k}
\del_{x'}\!\left\{\left[(R_{jk}^\ms{T})^\la_{\;\;\mu}n_k^{\dr;\mu}\right](x',t_\epsilon)\Theta(x-x')\right\}.
\label{g's eq}\end{equation}  
As seen from the first identity in Eq.s\,\eqref{g0} and Eq.\,\eqref{saddle eq.1}, this is equivalent to the following equality which correspond to Eq.\,(165) in \cite{BMFT 1}:
\begin{align}
n_j h\left(u_j^{(\la)}(x,t_\epsilon)\right)
&=n_j h_0\left(u_j^{(\la)}(x,t_\epsilon)\right)  \nn\\
&\quad+\frac{c'}{N}\sum_{k}\!
\del_{x'}\!\left\{\!\left[(R_{jk}^\ms{T})^\la_{\;\;\mu}n_k^{\dr;\mu}\!\right]\!\left(u_k^{(\mu)}(x',t_\epsilon)\!\right)\Theta\!\left(u_k^{(\mu)}(x,t_\epsilon)-u_k^{(\mu)}(x',t_\epsilon)\!\right)\!\right\}.
\label{165}\end{align}
However, using the dressed function formulae \eqref{dff} and  
\begin{align}
\del_p f_j^{\dr;\la}(x,t)
&=\sum_k\del_p(R^{-\ms{T}}_{jk})^\la_{\;\,\mu}(x,t)f_k^{(\mu)}  \nn\\
&=-\sum_{k,l} \left[(T^\dr_{jk})^\la_{\;\,\mu}\varsigma_k\left\{\del_p\vartheta_k^{(\mu)}\right\}(R^{-\ms{T}}_{kl})^\mu_{\;\,\nu}\right](x,t)f_l^{(\nu)}  \nn\\
&=-\sum_k \left[(T^\dr_{jk})^\la_{\;\,\mu}\varsigma_k\left\{\del_p\vartheta_k^{(\mu)}\right\}f_k^{\dr;\mu}\right](x,t)
\end{align}
with $p$ given in Eq.\,\eqref{dff}, one can bring Eq.\,\eqref{g's eq} back to Eq.\,\eqref{g's sol} as 
\begin{align}
n_j g^{(\la)}_j (x,t_\epsilon)
&=-N^{-1}c'n_j\delta(x-x')  \nn\\
&\quad
    -N^{-1}c'\left\{\del_{x'}n^{\dr;\la}_j(x',t_\epsilon) + \sum_{k} (T_{jk})^\la_{\;\;\mu}\varsigma_k \del_{x'}\!
     \left[\vartheta_k^{(\mu)}n^{\dr;\mu}_k\right]\!(x',t_\epsilon)\right\}\Theta(x-x')  \nn\\
&=-N^{-1}c'n_j\delta(x-x').  
\label{reform}\end{align}

Dressing the derivatives with respect to $c'$ of the both sides of Eq.\,\eqref{g's sol} and using the identity for dressed charges \eqref{eta sol.}, we have
\begin{equation}
\sum_k n_k \!\left[(R^{-\ms{T}}_{jk})^\la_{\;\,\mu} \del_{c'} g\right]\!\left(u_k^{(\mu)}\!(x,t_\epsilon)\right)
=
-N^{-1}n_j^{\dr;\la}\!\left(u_j^{(\la)}(x',t_\epsilon)\right)\delta(x-x').
\label{g's dress}\end{equation}
Inserting this into the dressed charge \eqref{3rd factor}, we obtain Eq.\,\eqref{version} at $c'=0$ in the limit $\epsilon\to0$ as 
\begin{align}
\left.\del_{c'}\ln\eta_j^{(\la)}(x,t')\right|_{c'=0}  
&=
2\beta N^{-1}\left\{n_j^{\dr;\la}({u'}^\la_j)_0\,\delta(x-x') + \mathfrak{S}_j^{(\la)}(x,t')_0  \right\}  \nn  
\end{align}
and
\begin{equation}
\mathfrak{S}_j^{(\la)}(x,t)_0  
=\frac{N}{2\beta} \frac{\del \ln\eta_j^{(\la)}(u_j^\la)_0}{\del u_j^\la} \left.\del_{c'}u_j^{(\la)}(x,t)\right|_{c'=0}.
\label{A.4}\end{equation}
On the other hand, differentiating Eq.\,\eqref{characteristics} with respect to $c'$ yields
\begin{align}
&\left.\del_{c'}u_j^{(\la)}(x,t)\right|_{c'=0}  \nn\\
&=\frac{1}{\rho_j^{\ms{t};\la}(u_j^\la)_0}
 \!\left\{
 \int_{-1/2}^x \!dy\, \del_{c'} \rho_j^{\ms{t};\la}\left(u_j^{(\la)}(y,t)\right)
 -\left.\int_{-1/2}^{u_j^\la}\!dy\,\del_{c'} \rho_j^{\ms{t};\la}(y)
 \right\}\right|_{c'=0}  \nn\\
&=\frac{1}{\rho_j^{\ms{t};\la}(u_j^\la)_0}
\sum_k\Biggl\{
\int_{-1/2}^x dy\,\varsigma_j\!\left[(T^\dr_{jk})^\la_{\;\,\mu} \Gamma_k^{(\mu)} \right](y,t)_0
\left.\del_{c'}\ln\eta_k^{(\mu)}(y,t)\right|_{c'=0} \nn\\ 
&\hspace{4.5cm}
-\int_{-1/2}^{u_j^\la}dy\,\varsigma_j \!\left[(T^\dr_{jk})^\la_{\;\,\mu} \Gamma_k^{(\mu)}\right]\!(y)_0 
 \left.\del_{c'}\ln\eta_k^{(\mu)}(y)\right|_{c'=0}
\Biggr\}.
\end{align}
We insert Eq.s \eqref{version} into this equation at $t=t'$ and then the resultant $\del_{c'}u_j^{(\la)}(x,t')|_{c'=0}$ into \eqref{A.4} at $t=t'$, which leads to
\begin{align}
\mathfrak{S}_j^{(\la)}(x,t')_0
=\mathfrak{S}_{0;j}^{(\la)}(x,t')_0  +\varsigma_j\mathcal{W}_j^{(\la)}&\!\left(u_j^{(\la)}(x,t')_0\right)_0 \sum_k\Biggl\{\int_{-1/2}^x \!dy\, 
    \!\left[(T^\dr_{jk})^\la_{\;\,\mu} \Gamma_k^{(\mu)} \mathfrak{S}_k^{(\mu)}\right]\!(y,t')_0  \nn\\
&\quad\quad-\int_{-1/2}^{u_j^{(\la)}(x,t')_0}dy \,\!\left[(T^\dr_{jk})^\la_{\;\,\mu} \Gamma_k^{(\mu)}  
                     \mathfrak{S}_k^{(\mu)}\right]\!(y)_0\Biggr\}, 
\label{d. charge}\end{align}
where $\mathcal{W}_j^{(\la)}(x)_0$ are given by \eqref{version 2} and $\mathfrak{S}_{0;j}^{(\la)}(x,t')_0$ are 
given by \eqref{d. term} with the initial value $\mathfrak{S}_{0;j}^{(\la)}(x)_0=0$, which means that there is no long-range spin correlation at the initial time. Thus, Eq. \eqref{d. charge} is reduced to Eq. \eqref{version 1}.

\end{appendix}


\begin{thebibliography}{9}
\bibitem{BMFT 1}
 B. Doyon, G. Perfetto, T. Sasamoto and T. Yoshimura,
“Ballistic macroscopic fluctuation theory",
SciPost Phys. \textbf{15}, 136 (2023).

\bibitem{BMFT 2}
B. Doyon, G. Perfetto, T. Sasamoto and T. Yoshimura,
“Emergence of Hydrodynamic Spatial Long-Range Correlations in Nonequilibrium Many-Body Systems",
Phys. Rev. Lett. \textbf{131}, 027101 (2023). 


\bibitem{KSS}
A. Kuniba, K. Sakai and J. Suzuki, 
“Continued fraction TBA and functional relations in XXZ model at root of unity”,
Nucl. Phys. B \textbf{525}, 597 (1998).

\bibitem{Doyon25(2)}
B. Doyon,
“NONLINEAR PROJECTION FOR BALLISTIC CORRELATION FUNCTIONS: A FORMULA IN TERMS OF MINIMAL CONNECTED COVERS”,
arXiv:2506.05266v2 [cond-mat.stat-mech] (2025)

\bibitem{AS}
S. Ae and K. Sakai,
“Spin Drude weight for the integrable XXZ chain with arbitrary spin”,
J. Stat. Mech. (2024) 033104.

\bibitem{Hubner etc.}
F. Hübner, L. Biagetti, J. D. Nardis and B. Doyon,
“Diffusive Hydrodynamics from Long-Range Correlations",
Phys. Rev. Lett. \textbf{134}, 187101 (2025).

\bibitem{Doyon lec.}
B. Doyon,
“Lecture notes on generalised hydrodynamics”,
SciPost Phys. Lect. Notes 18 (2020).

\bibitem{Ae 2024}
S. Ae,
“Infinite temperature spin dc conductivity of the spin-1/2 XXZ chain”,
J. Phys. A: Math. Theor. \textbf{57}, 445204 (2024).

\bibitem{Yoshimura & Krajnik}
T. Yoshimura and Ž. Krajnik,
“Anomalous current fluctuations from Euler hydrodynamics",
Phys. Rev. E \textbf{111}, 024141 (2025).

\bibitem{SCCA}
K. Klobas, M. Medenjak and T. Prosen,
“Exactly solvable deterministic lattice model of crossover between ballistic and diffusive transport”,
J. Stat. Mech. (2018) 123202.

\bibitem{YKBI 26}
T. Yoshimura, {\"z}. Krajnik, A. Bastianello and E. Ilievski
“Anomalous hydrodynamic fluctuations in the quantum XXZ spin chain”,
arXiv:2602.24242v3 [cond-mat.stat-mech] (2026).

\bibitem{TS 72}
M. Takahashi and M. Suzuki, 
“One-Dimensional Anisotropic Heisenberg Model at Finite Temperatures”,
Prog. Theor. Phys. \textbf{48}, 2187 (1972). 

\bibitem{Takahashi 99}
M. Takahashi,
“Thermodynamics of One-dimensional Solvable Models”, 
Cambridge University Press (1999).


\bibitem{ND 16}
B. Bertini, M. Collura, J. D. Nardis and M. Fagotti,
“Transport in Out-of-Equilibrium XXZ Chains: Exact Profiles of Charges and Currents”,
Phys. Rev. Lett. \textbf{117}, 207201 (2016).

\bibitem{Doyon 16}
O. A. Castro-Alvaredo, B. Doyon and T. Yoshimura,
“Emergent Hydrodynamics in Integrable Quantum Systems out of Equilibrium”,
Phys. Rev. X \textbf{6}, 041065 (2016).

\bibitem{BVKM}
V. B. Bulchandani, R. Vasseur, C. Karrasch and J. E. Moore,
“Bethe-Boltzmann hydrodynamics and spin transport in the XXZ chain”,
Phys. Rev. B \textbf{97}, 045407 (2018).

\bibitem{Korepin}
V. E. Korepin, N. M. Bogoliubov and A. G. Izergin,
“Quantum Inverse Scattering Method and Correlation Functions”,
Cambridge University Press (1993).

\bibitem{force}
B. Doyon and T. Yoshimura,
“A note on generalized hydrodynamics: inhomogeneous fields and other concepts”,
SciPost Phys. \textbf{2}, 014 (2017).

\bibitem{DSY curve}
B. Doyon, H. Spohn and T. Yoshimura,
“A geometric viewpoint on generalized hydrodynamics”,
Nucl. Phys. B \textbf{926}, 570 (2018).

\bibitem{Eggert} 
S. Eggert, I. Affleck and M. Takahashi,
“Susceptibility of the spin 1/2 Heisenberg antiferromagnetic chain"
Phys. Rev. Lett. \textbf{73}, 332 (1994)

\bibitem{Kubo}
R. Kubo, 
“Statistical-Mechanical Theory of Irreversible Processes. I. General Theory and
Simple Applications to Magnetic and Conduction Problems”, 
J. Phys. Soc. Jpn. \textbf{12}, 570 (1957).

\bibitem{Zotos}
X. Zotos,
“Finite Temperature Drude Weight of the One-Dimensional Spin- 
1/2 Heisenberg Model”,
Phys. Rev. Lett. \textbf{82}, 1764 (1999).

\bibitem{Klumper}
A. Urichuk, Y. {\"O}z, A. Kl{\"u}mper and J. Sirker, 
“The spin Drude weight of the XXZ chain and generalized hydrodynamics”, 
SciPost Phys. \textbf{6}, 005 (2019).

\bibitem{Kirillov}
A. Kirillov and N. Reshetikhin,
“Classification of the string solutions of Bethe equations in an XXZ model of arbitrary spin”,
Zap. Nauch. Semin. LOMI \textbf{146}, 31-46 (1985).

\bibitem{Kirillov2}
A. Kirillov and N. Reshetikhin,
“Classification of the string solutions of {Bethe} equations in an {XXZ} model of arbitrary spin”,
J. Sov. Math. \textbf{40}, 22-35 (1988).

\bibitem{form factor}
J. D. Nardis and M. Panfil,
“Particle-hole pairs and density–density correlations in the Lieb–Liniger model",
J. Stat. Mech. (2018) 033102.

\bibitem{ND diffusion}
J. D. Nardis, D. Bernard and B. Doyon,
“Diffusion in generalized hydrodynamics and quasiparticle scattering”,
SciPost Phys. \textbf{6}, 049 (2019).

\bibitem{Takahashi XXX}
M. Takahashi,
“One-Dimensional Heisenberg Model at Finite Temperatures”,
Prog. Theor. Phys. \textbf{46}, 401 (1971). 

\bibitem{ND diffusion (0)}
E. Ilievski, J. D. Nardis, M. Medenjak and T. Prosen,
“Superdiffusion in One-Dimensional Quantum Lattice Models”,
Phys. Rev. Lett. \textbf{121}, 230602 (2018).

\bibitem{Gopalakrishnan2}
S. Gopalakrishnan and R. Vasseur, 
“Kinetic theory of spin diffusion and superdiffusion in XXZ spin chains”, 
Phys. Rev. Letters \textbf{122}, 127202 (2019).

\bibitem{KPZ}
M. Kardar, G. Parisi and Y. C. Zhang,
“Dynamic Scaling of Growing Interfaces”,
Phys. Rev. Lett. \textbf{56}, 889 (1986).

\bibitem{KPZ-XXX}
M. Ljubotina, M. {\v{Z}}nidari{\v{c}}, and T. Prosen,
“Kardar-Parisi-Zhang physics in the quantum Heisenberg magnet”,
Phys.  Rev. Lett. \textbf{122}, 210602 (2019).

\bibitem{KPZ no XXX}
{\v{Z}}. Krajnik, E. Ilievski and T. Prosen,
“Absence of Normal Fluctuations in an Integrable Magnet”,
Phys. Rev. Lett. \textbf{128}, 090604 (2022).

\bibitem{KPZ no XXX 2}
E. Rosenberg et al.,
“Dynamics of magnetization at infinite temperature in a Heisenberg spin chain”,
Science \textbf{384}, 48-53 (2024).


\end{thebibliography}
\end{document}